\documentclass[showpacs,prl,twocolumn,floatfix,reprint]{revtex4}
\usepackage{graphicx,amssymb,amsmath}
\newcommand{\E}{\mathrm{E}}
\newcommand{\Var}{\mathrm{Var}}
\newcommand{\Cov}{\mathrm{Cov}}

\begin{document}

\title{Patterns of residential segregation}

\author{R\'emi Louf}
\email{r.louf@ucl.ac.uk}
\affiliation{Institut de Physique Th\'{e}orique, CEA, CNRS-URA 2306, F-91191, 
Gif-sur-Yvette, France}
\affiliation{Centre for Advanced Spatial Analysis (CASA), University College
London, 90 Tottenham Court Road, W1T 4TJ, London, UK}

\author{Marc Barthelemy}
\email{marc.barthelemy@cea.fr}
\affiliation{Institut de Physique Th\'{e}orique, CEA, CNRS-URA 2306, F-91191, 
Gif-sur-Yvette, France}
\affiliation{CAMS (CNRS/EHESS) 190-198, avenue de France, Paris, France}

\begin{abstract}
The spatial distribution of income shapes the structure and
organisation of cities and its understanding has broad societal
implications. Despite an abundant literature, many issues remain
unclear. In particular, all definitions of segregation are implicitely
tied to a single indicator, usually rely on an ambiguous definition of
income classes, without any consensus on how to define neighbourhoods
and to deal with the polycentric organization of large cities. In this
paper, we address all these questions within a unique conceptual
framework. We avoid the challenge of providing a direct definition of
segregation and instead start from a definition of what segregation is
not. This naturally leads to the measure of representation that is
able to identify locations where categories are over- or
underrepresented. From there, we provide a new measure of exposure
that discriminates between situations where categories co-locate or
repel one another. We then use this feature to provide an unambiguous,
parameter-free method to find meaningful breaks in the income
distribution, thus defining classes. Applied to the 2014 American
Community Survey, we find 3 emerging classes -- low, middle and higher
income -- out of the original 16 income categories. The higher-income
households are proportionally more present in larger cities, while
lower-income households are not, invalidating the idea of an increased
social polarisation. Finally, using the density -- and not the
distance to a center which is meaningless in polycentric cities -- we
find that the richer class is overrepresented in high density zones,
especially for larger cities. This suggests that density is a relevant
factor for understanding the income structure of cities and might
explain some of the differences observed between US and European
cities.
\end{abstract}

\maketitle

\section*{Introduction}

Challenges posed by the constantly growing urbanisation are
complex and difficult to handle. They range from the increasing
dependence on energy, to serious environmental and sustainability
issues, and socio-spatial inequalities \cite{UN}. In particular, we
observe the appearance of socially homogeneous zones and dynamical
phenomena such as urban decay \cite{Jacobs:1961,Andersen:2002} and gentrification
\cite{Atkinson:2004} that reinforce the heterogeneity of the spatial
distribution of social classes in cities. Such a segregation
-- characterized by an important social differentiation of the urban space -- 
has significant social, economic \cite{Massey:1993} and even health
costs \cite{Lobmayer:2002} which justify the attention it has attracted in
academic studies over the past century. Despite the abundant
literature in sociology and economics, however, there is no consensus
on the adequate way to quantify and describe patterns of segregation.
In particular, the identification of
neighbourhoods where the different groups gather is still in its infancy.

As stated many times, and at different periods in the sociology
literature~\cite{Duncan:1955,James:1982,Massey:1988,Reardon:2002}, the study of
segregation is cursed by its intuitive appeal. The perceived familiarity with the concept
favours what Duncan and Duncan~\cite{Duncan:1955} called `naive operationalism':
the tendency to force a sociological interpretation on measures that are at odds
with the conceptual understanding of segregation. As a matter of fact,
segregation is a complex notion, and the literature distinguishes several
conceptually different dimensions. Massey~\cite{Massey:1988} first proposed a
list of $5$ dimensions (and related existing measures), which was recently
reduced to $4$ by Reardon~\cite{Reardon:2004}. (i) {\it exposure} which measures
the extent to which different populations share the same residential areas; (ii)
the {\it evenness} (and {\it clustering}) to which extent populations are evenly
spread in the metropolitan area; (iii) {\it concentration} to which extent
populations concentrate in the areal units they occupy; and (iv) {\it
centralization} to which extent populations concentrate in the center of the
city.

We identify several problems with this picture. The first -- fundamental --
issue lies in the lack of a general conceptual framework in which all existing
measures can be interpreted.  Instead, we have a patchwork of seemingly
unrelated measures that are labelled with either of the aforementioned
dimensions. Although segregation can indeed manifest itself in different ways,
it is relatively straightforward to define what is \emph{not} segregation: a
spatial distribution of different categories that is undistinguishable from a
uniform random situation (with the same percentages of different categories).
Therefore, we can define segregation as {\it any pattern in the spatial
distribution of categories that deviates significantly from a random
distribution}~\cite{Winship:1977}. The different dimensions
of~\cite{Massey:1988,Reardon:2004} then correspond to particular aspects of how
a multi-dimensional pattern can deviate from its randomized counterpart. The
measures we propose here are all rooted in this general definition of
segregation.

The other issues are technical in nature. First, several difficulties
are tied to the existence of many categories in the underlying
data. Historically, measurements of racial segregation were limited to
measures between $2$ population groups. However, most measures
generalise poorly to a situation with many groups, and the others do
not necessarily have a clear
interpretation~\cite{Reardon:2002}. Worse, in the case of groups based
on a continuum (such as income), the thresholds chosen to define
classes are usually arbitrary~\cite{Jargowsky:1996}. We propose in the
following to solve this issue by defining classes in a unambiguous and
non-arbitrary way through their pattern of spatial
interaction. Applied to the distribution of income categories in US
cities, we find $3$ emergent categories, which are naturally
interpreted as the lower-, middle- and higher-income classes. Second,
most authors systematically design a single index of segregation for
territories that can be very large, up to thousands of square
kilometers~\cite{Apparicio:2000}. In order to mitigate segregation, a
more local, spatial information is however needed: local authorities
need to locate where the poorest and richest concentrate if they want
to design efficient policies to curb, or compensate for the existing
segregation. In other words, we need to provide a clear {\it spatial}
information on the pattern of segregation. Previous
studies~\cite{Ellis:2004,Lobo:2007,Wong:2010, Sharma:2012} were
interested in the characterisation of intra-urban segregation
patterns, but they suffer from the limitations of the indicators they
use. In particular, the values they map come with no indication as to
when a high value of the index indicates high segregation levels. As a
result, the maps are not necessarily easy to read. Furthermore, all
the descriptions are cartographic in nature and while maps are a
powerful way to highlight patterns, we would like to provide further,
quantitative, information about the spatial distribution that goes
beyond cartographic representation.

The lack of a clear characterization of the spatial distribution of individuals is 
not tied to the problem of segregation in particular, but pertains to the field
of spatial statistics~\cite{Ripley:1981,Cressie:1993,Chun:2013}. Many studies avoided this spatial
problem by assuming implicitely that cities are monocentric and circular, and rely on either an
arbitrary definition of the city center boundaries, or on indices computed as a
function of the distance to the center (whatever this may be). However, most if
not all cities are anisotropic, and the large ones, polycentric (see~\cite{Louf:2013}
and references therein). Many empirical studies and models in economics aim to
explain the difference between central cities and suburbs \cite{Glaeser:2008,
Brueckner:1999}. Yet, the sole stylized fact upon which they rely -- city centers
tend to be poorer than suburbs (in the US) -- lacks a solid empirical
basis.

In the first part of the paper, we define a null model -- the
unsegregated city -- and define the representation, a measure that
identifies significant local departures from this null case. We
further introduce a measure of exposure that allows us to quantify the
extent to which the different categories attract or repel one
another. This exposure is the starting point for the non-parametric
identification of the different social classes.  In the second part,
we define neighbourhoods by clustering adjacent areal units where
classes are overrepresented and show that there an increased spatial
isolation of classes as population size of cities grows. We also show
that larger cities are richer in the sense that the wealthiest
households tend to be overrepresented and the low-income
underrepresented in large cities. Finally, we discuss how density is
connected to the spatial distribution of income, and how to go beyond
the traditional picture of a poor center and rich suburbs.

We focus here on the income distribution, using the data for the 2014 Core-Based
Statistical areas. However, the methods presented in this paper are very
general, and can be applied~\cite{github:marble} to different geographical
levels, to an arbitrary number of population categories, and to different
variables such as ethnicity, education level, etc.

\section{The importance of a null model}
 \label{sub:null_model}

Most studies exploring the question of spatial segregation define
measures before comparing their value for different cities. Knowing
that two quantities are different is however not enough: we also have
to know whether this difference is significant. In order to assess
the significance of a result, we have to compare it to what is
obtained for a reasonable null model.

\subsection{Definitions}

We assume that we have $T$ areal units dividing the city and that
individuals can belong to different categories. The elementary
quantity is $n_\alpha(t)$ which represents the number of individuals
of category $\alpha$ in the unit $t$. The total number of individuals
belonging to a category $\alpha$ is $N_\alpha$ and the total number of
individuals in the city is given by $N=\sum_\alpha N_\alpha$. 

In the context of residential segregation, a natural null model is the
\emph{unsegregated city}, where all households are distributed at
random in the city with the constraints that
\begin{itemize}
    \item The total number $n(t)$ of households living in the areal unit $t$ is
	    fixed (from data).
    \item The numbers $N_\alpha$ are given by the data; 
\end{itemize}
The problem of finding the numbers
$\{ n_\alpha(1), \dots, n_\alpha(T) \}$ in this unsegregated
city is reminiscent of the traditional occupancy problem in
combinatorics~\cite{Feller:1950}. If we assume that for all categories
$\alpha$, we have $n_\alpha (t)\ll n(t)$, they are then distributed
according to the multinomial denoted by
$f \left( n_\alpha(1), \dots, n_\alpha(T) \right)$, and the number of
people of category $\alpha$ in the areal unit $t$ is distributed
according to a binomial
distribution.  Therefore, in an unsegregated city, we have
\begin{align}
    \E \left[ n_\alpha(t) \right] &= N_\alpha\,\frac{n(t)}{N} \\
    \Var \left[ n_\alpha(t) \right] &= N_\alpha\,\frac{n(t)}{N} \left( 1 - \frac{n(t)}{N}  \right) 
\end{align}

The fundamental quantity we will use
in the following is the {\em representation} of a category $\alpha$ in the areal unit $t$,  defined as
\begin{equation}
    r_\alpha(t) = \frac{n_\alpha(t) / n(t)}{N_\alpha / N} = \frac{n_\alpha(t) /
    N_\alpha}{n(t)/N} 
\label{eq:repre}
\end{equation}
The representation thus compares the relative population $\alpha$ in
the areal unit $t$ to the value that is expected in an unsegregated city where
individuals choose their location at random. Or, equivalently, the
representation compares the proportion of individuals $\alpha$ in the unit $t$
to their proportion in the city as a whole.

In metropolitan areas, $N_\alpha$ is large compared to $1$, and the
distribution of the $n_\alpha(t)$ can be approximated by a Gaussian
with the same mean and variance. Therefore we have in the unsegregated case

\begin{align}
    \begin{split}
	\E \left[ r_\alpha(t) \right] &= 1 \\
        \Var \left[ r_\alpha(t) \right] &= \sigma_\alpha(t)^2 = \frac{1}{N_\alpha} \left(
    \frac{N}{n(t)} - 1  \right) 
    \end{split}
\end{align}

An important merit of the representation is the possibility to define rigorously
the notion of {\it over-}representation and {\it under-}representation of a
category $\alpha$ in a geographical area. A category $\alpha$ is
overrepresented (with a $99\%$ confidence) in the geographical area $t$ if
$r_\alpha(t) > 1 + 2.57\,\sigma_\alpha(t)$.  A category $\alpha$ is
underrepresented (with a $99\%$ confidence) in the geographical area $t$ if
$r_\alpha(t) < 1 - 2.57\,\sigma_\alpha(t)$. If the value $r_\alpha(t)$ falls in
between the two previous limits, the representation of the category $\alpha$
is not statistically different (at this confidence level) from what
would be obtained if individuals were distributed at random. Existing measures
output levels of segregation (typically a number between 0 and 1) but do not indicate 
 whether these levels are \emph{abnormally} high. To this respect, the representation is a
significant improvement over previous measures. 

Note that the above null model is reminiscent of the `counterfactuals'
used in the empirical literature on agglomeration
economies~\cite{Duranton:2005, Marcon:2009, Billings:2012}. Also, the
expression of the representation (Eq.  \ref{eq:repre}) is very similar
to the formula used in economics to compute comparative
advantages~\cite{Balassa:1965}, or to the localisation quotient used
in various contexts~\cite{Apparicio:2000, Schwabe:2011}.  To our
knowledge, however, this formula has never been justified by a null
model in the context of residential location. The representation
allows to assess the significance of the deviation of population
distributions from the unsegregated city. As we will show below, it is
also the building block for measuring the level of repulsion or
attraction between categories allowing us to uncover the different
classes and to identify the neighbourhoods where the different
categories concentrate. Last, but not least, the representation
defined here does not depend on the category structure at the city
scale, but only on the spatial repartition of individuals belonging to
each category. This is essential in order to be able to compare
different cities where the group compositions -- or inequality --
might differ. Inequality and segregation are indeed two separate
concepts, and the way they are measured should be distinct from one
another.

Finally, we would like to mention that using the uniform distribution as a null
model can have implications broader than the study of residential segregation.
Indeed, from a very abstract perspective, the study of residential segregation
is the study of labelled objects in space. The methods presented here can
therefore be applied to the study of the distribution of any object in
space. In particular, it can be used to identify the locations in a territory where
populations with different characteristics (not necessarily
socio-economic) concentrate.

\subsection{Attraction and repulsion of categories}
    \label{sub:attraction_and_repulsion_of_populations}

Another shortcoming of the literature about segregation is the lack of indicator
to quantify to what extent different populations attract or repel one another.
Such a measure of attraction or repulsion is however important to
understand the dynamics and scale (intensity of attraction/repulsion) of residential
segregation.

Our indicator is inspired by the M-value first introduced by Marcon
and Puech in the economics literature to measure the concentration of
industries~\cite{Marcon:2009} and used as a measure of interaction
between retail store categories in~\cite{Jensen:2006}.  These authors
were interested in measuring the geographic concentration of different
type of industries. While previous measures (such as Ripley's K-value)
allow to identify departures from a random (Poisson) distribution, the
M-value's interest resides in the possibility to evaluate different
industries' tendency to co-locate. The idea, in the context of
segregation is simple: we consider two categories $\alpha$ and $\beta$
and we would like to measure to which extent they are co-located in
the same areal unit. To quantify the tendency of households to
co-locate, we measure the representation of the category $\beta$ as
witnessed on average by individuals in category $\alpha$, and
obtain the following quantity $E_{\alpha\beta}$
\begin{equation} 
    E_{\alpha \beta} = \frac{1}{N_\alpha} \sum_{t=1}^{T} n_\alpha(t)\,r_\beta(t) 
\end{equation}
Although it is not obvious with this formulation, this measure is symmetric:
$E_{\alpha \beta} = E_{\beta \alpha}$
(see Supplementary Information below. Effectively, this `E-value' in this context is a measure of exposure, according to the typology of
segregation measures proposed in~\cite{Massey:1988}. However, unlike the other
measures of exposure found in the literature~\cite{Bell:1954}, we are able to
distinguish between situations where categories attract ($E>1$)
or repel ($E<1$) one another.  In the case of an unsegregated city, every household in
$\alpha$ sees on average $r_\beta = 1$ and we have $E_{\alpha \beta} = 1$. If
populations $\alpha$ and $\beta$ attract each other, that is if they tend to be
overrepresented in the same areal units, every household $\alpha$ sees $r_\beta
> 1$ and we have $E_{\alpha \beta} > 1$ at the city scale. On the other hand, if
they repel each other, every household $\alpha$ sees $r_\beta < 1$ and we have
$E_{\alpha \beta} <1$ at the city scale. The minimum of the exposure for two
classes $\alpha$ and $\beta$ is obtained when these two categories are never
present together in the same areal unit and then
\begin{equation}
    E_{\alpha\,\beta}^{min} = 0 
\end{equation}
The maximum is obtained when the two classes are alone in the
system (see Supplementary Information below for more details) and in this case we get
\begin{equation}
    E_{\alpha\,\beta}^{max} = \frac{N^2}{4N_\alpha N_\beta}
\end{equation}
In the case $\alpha = \beta$, the previous measure represents the
`isolation' defined as
\begin{equation}
    I_\alpha = \frac{1}{N_\alpha}\sum_{t=1}^{t} n_\alpha(t)\,r_\alpha(t)
\end{equation}
and measures to which extent individuals from the same category
interact which each other. In the unsegregated city, where individuals are
indifferent to others when chosing their residence, we have
$I_\alpha^{min}=1$. In contrast, in the extreme situation where
individuals belonging to the class $\alpha$ live isolated from the
others, the isolation reaches its maximum value
\begin{equation}
    I_\alpha^{max} = \frac{N}{N_\alpha}
\end{equation}
Of course, in order to discuss the significance of the values of exposure and
isolation, one needs to compute the variance of the exposure in the unsegregated
situation defined earlier. The calculations for the variance as well as for the
extrema are presented in the Supplementary Information below. 

Finally, we note that co-location is not necessarily synonymous with
interaction, as pointed out by Chamboredon~\cite{Chamboredon:1970},
and we should rigorously talk about \emph{potential}
interactions. Nevertheless, in the absence of large scale data about
direct interactions between individuals, co-location is the best proxy
available.

\section{Emergent social classes}
\label{sub:uncovering_the_social_stratification_of_cities}

\subsection{Defining classes}
\label{sub:}

Studies that focus on the definition of a single segregation index for cities as
a whole can avoid the problem of defining classes, either by measuring the
between-neighbourhood variation of the average income (examples are the standard
deviation of incomes~\cite{Jargowsky:1996}, the variance of logged
incomes~\cite{Ioannides:2004} and Jargowsky's
Neighbourhood Sorting Index~\cite{Jargowsky:1996}), or by integrating over the
entire income distribution (for instance the rank-order information theory index defined
in~\cite{Reardon:2006}). However, when they investigate the behaviour of households
with different income and their spatial distribution, studies of segregation
must be rooted in a particular definition of categories (or
classes). Unfortunately, there is no consensus in the
literature about how to separate households in different classes according to
their income, and studies generally rely on more or less arbitrary
divisions~\cite{Massey:1996, Massey:2003, Jenkins:2006}.

While in some particular cases grouping the original categories in pre-defined
classes is justified, most authors do so for mere convenience reasons. However,
as some sociologists have already pointed out~\cite{Emirbayer:1997}, imposing
the existence of absolute, artificial entities is necessarily going to bias our
reading of the data. Furthermore, in the absence of recognized standards,
different authors will likely have different definitions of classes, making the
comparisons between different results in the literature difficult. 

From a theoretical point of view, entities such as social classes do not have an
existence of their own. Grouping the individuals into arbitrary classes when
studying segregation is thus a logical fallacy: it amounts to imposing a class
structure on the society before assessing the existence of this structure (which
manifests itself by the differentiated spatial repartition of individuals with
different income). Here, instead of imposing an arbitrary class
structure, we let the class structure emerge from the data themselves. Our
starting hypothesis is the following: {\it if there is such a thing as a social
stratification based on income, it should be reflected in the households'
behaviours}: households belonging to the same class should tend to live together,
while households belonging to different classes should tend to avoid one
another. In other words, we aim to define classes using the way they manifest
themselves through the spatial repartition of the different categories.

\subsection{Finding breaks in the income distribution}
\label{sub:the_method}

We choose as a starting point the finest income subdivision given by the US Census
Bureau ($16$ subdivisions) and compute the $16 \times 16$ matrix of $E_{\alpha
\beta}$ values for all cities. We then perform a hierarchical clustering on this
matrix, succesively aggregating the subdivisions with the highest $E_{\alpha
\beta}$ values. The process, that we implemented in the Python library
Marble~\cite{github:marble},  goes as follows:

\begin{enumerate}
    \item Check whether there exists a pair $\alpha$, $\beta$ such that
        $E_{\alpha \beta} > 1 + 10\,\sigma$ (i.e. two categories that attract
        one another with at least 99\% confidence according to the Chebyshev
        inequality). If not, stop the agregation and return the classes;
    \item If there is at least one couple satisfying (1), normalize all $E_{\alpha \beta}$ values by their
        respective maximum values. Find then the pair $\gamma$, $\beta$ whose
        normalized exposure is the maximum;
    \item Aggregate the two categories $\beta$ and $\gamma$;
    \item Repeat the process until it stops.
\end{enumerate}

In order to aggregate the categories at step 3, we need to compute the exposure 
between  $\delta = \beta \cup \gamma$ and any category $\alpha$, as well as its
variance. The corresponding calculations are presented in the Supplementary
Information (below). 

We stress that the obtained classification does not rely on any arbitrary
threshold. Indeed, we stop the aggregation process when the only classes left are
indifferent ($E_{\alpha \beta} = 1$ with $99\%$ confidence) or
repel each other ($E_{\alpha \beta} < 1$ with $99\%$ confidence).

\subsection{The US income structure}
\label{sub:social_structure_in_the_us}

\begin{figure*}
    \includegraphics[width=0.95\textwidth]{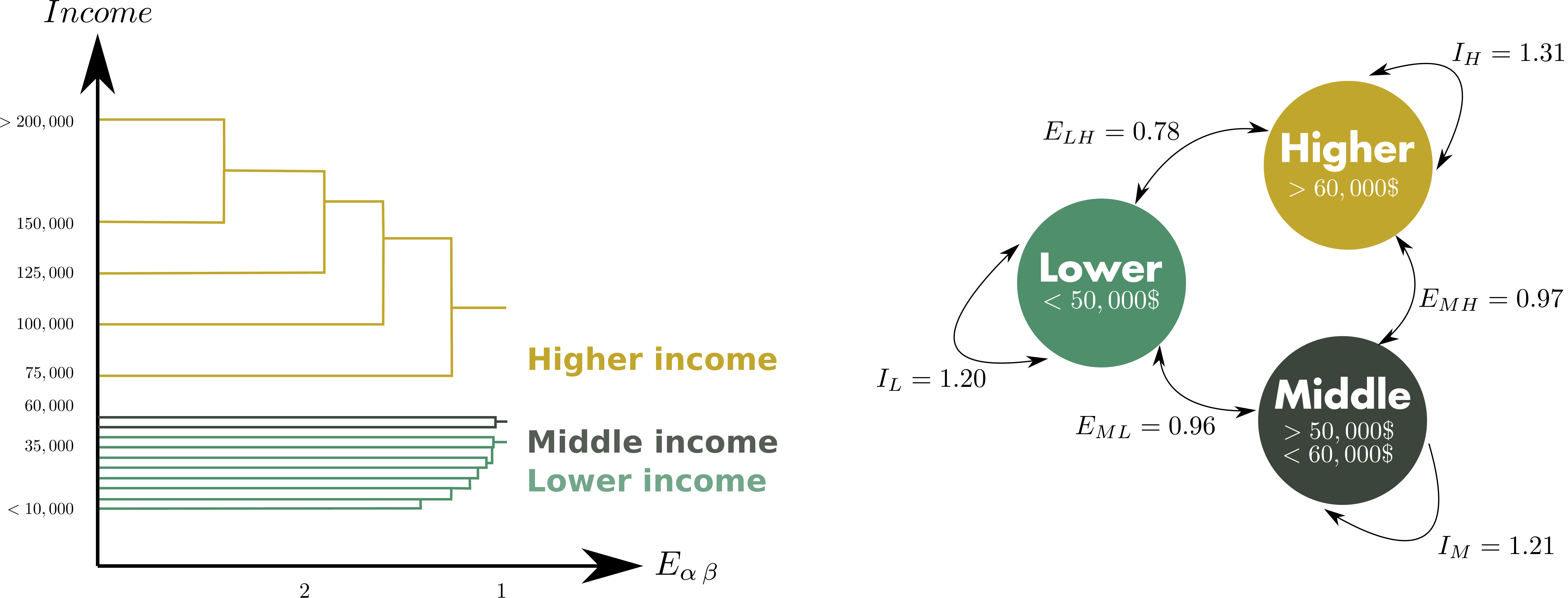}
    \caption{(Left) Alluvial diagram showing the successive aggregation of
    different income categories during the clustering process. The x-axis shows the
    value of the exposure at which each aggregation took place. The
    aggregation stops when there is no pair of category ($\alpha$, $\beta$) for which
    $E_{\alpha \beta} > 1$, that is when all classes are at best indifferent to one
    another (see Supplementary Information, section 2 for a more detailed
    description of the algorithm). 
    One can see on this diagram that the highest income categories tend to
    colocate more (higher values of $E_{\alpha \beta}$) than the lowest
    income categories. 
    (Right) The 3 classes that emerge from the clustering process.
    In the circles we indicate the range of income to which each class
    corresponds. The value next to the arrows correspond to the respective
    values of exposure and isolation. As the values of exposure show, the lower- and
    higher-income classes repel one another, while the middle-income
    class is indifferent to the other classes.  Furthermore, the higher-income
    class is a more coherent group than the middle-income and lower-income
    classes, as reflected by the values of the 
    isolation coefficient $I$.}
\label{fig:classes_alluvial}
\end{figure*}

Strikingly, the outcome of this method on US data is the emergence of 3 distinct classes
(Fig.~\ref{fig:classes_alluvial}):
the higher-income ($\sim 29\%$ of the US population) and the lower-income classes
($\sim 59\%$ of the US population) -- which repel each other strongly while being
respectively very coherent -- and a somewhat small middle-income class
($\sim 11\%$ of the population) that is relatively indifferent to the other classes. This
result implies that there is some truth in the conventional way of dividing
populations into $3$ income classes, and that what we casually perceive as the
social stratification in our cities actually emerges from the spatial
interaction of people. 

Our method has several advantages over a casual
definition: it is not arbitrary in the sense that it does not depend on a
tunable parameter (besides the significance threshold) and on who performs the analysis. Its origins are tractable,
and can be argued on a quantitative basis. Because it is quantitative, it allows
comparison of the stratification over different points in time, or between
different countries. It can also be compared to other class divisions that would
be obtained using a different medium for interaction, for instance mobile phone
communications~\cite{Eagle:2010}. 

In the following, we will systematically use the classes obtained with
this method.

\subsection{Larger cities are richer}
\label{sub:inter_urban}

\begin{figure}
    \centering
    \includegraphics[width=0.49\textwidth]{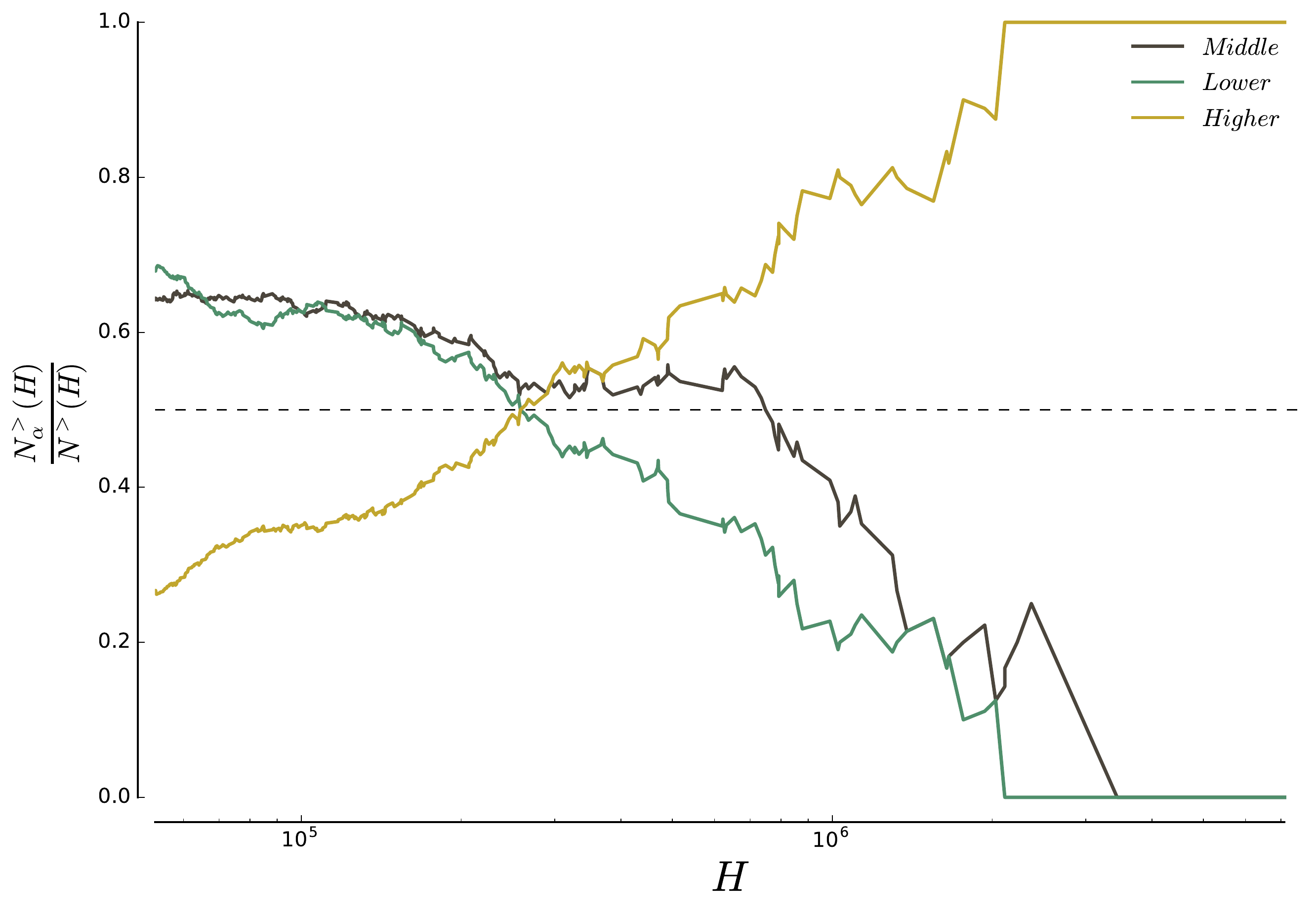}
    \caption{Proportion of cities in which the different classes are
    overrepresented, as a function of the total population of the city. One can
    clearly see that the larger the city, the more likely it is that the high-income class is
    overrepresented and the lower-income class underrepresented (compared to
    national levels). This proves that the different income classes are not
    homogeneously distributed across all cities in the country. There
    is also a clear influence of population size on the representation of the different
    classes at the city level.
\label{fig:inter-urban_representation}}
\end{figure}

At the scale of an entire country, segregation can manifest itself in the
unequal representation of the different income classes across the urban areas.
We plot on Fig.~\ref{fig:inter-urban_representation} the ratio 
$N_\alpha^{>}(H)/N^{>}(H)$ where $N^{>}(H)$ is the number of cities of
population greater than $H$, and $N_\alpha^{>}(H)$ the number of cities of
population greater than $H$ for which the class $\alpha$ is
overrepresented. A decreasing curve indicates that the class $\alpha$ tends to be
underrepresented in larger urban areas, while an increasing curve shows that the
category $\alpha$ tends to be overrepresented in larger urban areas (the
representation is here measured with respect to the total population
at the US level). These results challenge Sassen's thesis on social
polarization, according to which world (very large) cities host
proportionally more higher-income and lower-income individuals than smaller
cities~\cite{Sassen:1991}. If this thesis were correct, we should
observe an overrepresentation of both higher-income and lower-income households in larger
cities. Instead, as shown on Fig.~\ref{fig:inter-urban_representation},
higher-income households are overrepresented in larger cities, while
lower-income households tend to be underrepresented (see Supplementary
Information for a
detailed discussion). These results support the previous critique of the social
polarisation thesis by Hamnett~\cite{Hamnett:1994}.

\section{Characterizing spatial patterns}
\label{sec:the_spatial_pattern_of_segregation}

The representation measure introduced at the beginning of this article allows to
draw maps of overrepresentation and thus to identify the areas of the city
where categories are overrepresented. In the following, we propose to
characterise the spatial arrangement of these areas for the different
categories. 

\subsection{Poor center, rich suburbs?}
\label{sub:poor_center_rich_suburbs_}

\subsubsection{A density-based method}
\label{sec:A density-based method}

In many studies, the question of the spatial pattern of segregation is limited
to the study of the center versus suburbs and is usually adressed in two
different ways. 

In the first case, a central area is defined by arbitrary boundaries and
measures are performed at the scale of this central area and the rest
is labelled as `suburbs'. The issue with this approach is that the
conclusions depend on the chosen boundaries and there is no unique unambiguous
definition of the city center: while some consider the Central Business
District~\cite{Glaeser:2008}, others choose the urban
core (urbanized area) where the population density is higher. 

The second approach, in an attempt to get rid of arbitrary boundaries, consists
in plotting indicators of wealth as a function of distance to the
center~\cite{Glaeser:2008, Brueckner:2009}. This approach, inspired by the
monocentric and isotropic city of many economic studies such as the Von Thunen
or the Alonso-Muth-Mills model~\cite{Brueckner:1987}, has however a serious
flaw: cities are not isotropic and are spread unevenly in space, leading to very
irregular shapes~\cite{Makse:1995}. Representing any quantity versus the
distance to a center thus amounts to average over very different areas and in
polycentric cases (as it is the case for large cities \cite{Louf:2013}) is
necessarily misleading. As we show below, this method mixes together areas that
are otherwise very different.

We propose here a different approach that does not require to draw boundaries
between the center and the suburbs. In fact, it does not even require to define
and locate the `center' at all. In the case of a monocentric and isotropic city,
our method gives results similar to those given by the other measures. In the
more general case where cities are not necessarily monocentric neither
isotropic, our method allows to compare regions of equivalent densities.

\begin{figure*}
    \centering
    \includegraphics[width=0.95\textwidth]{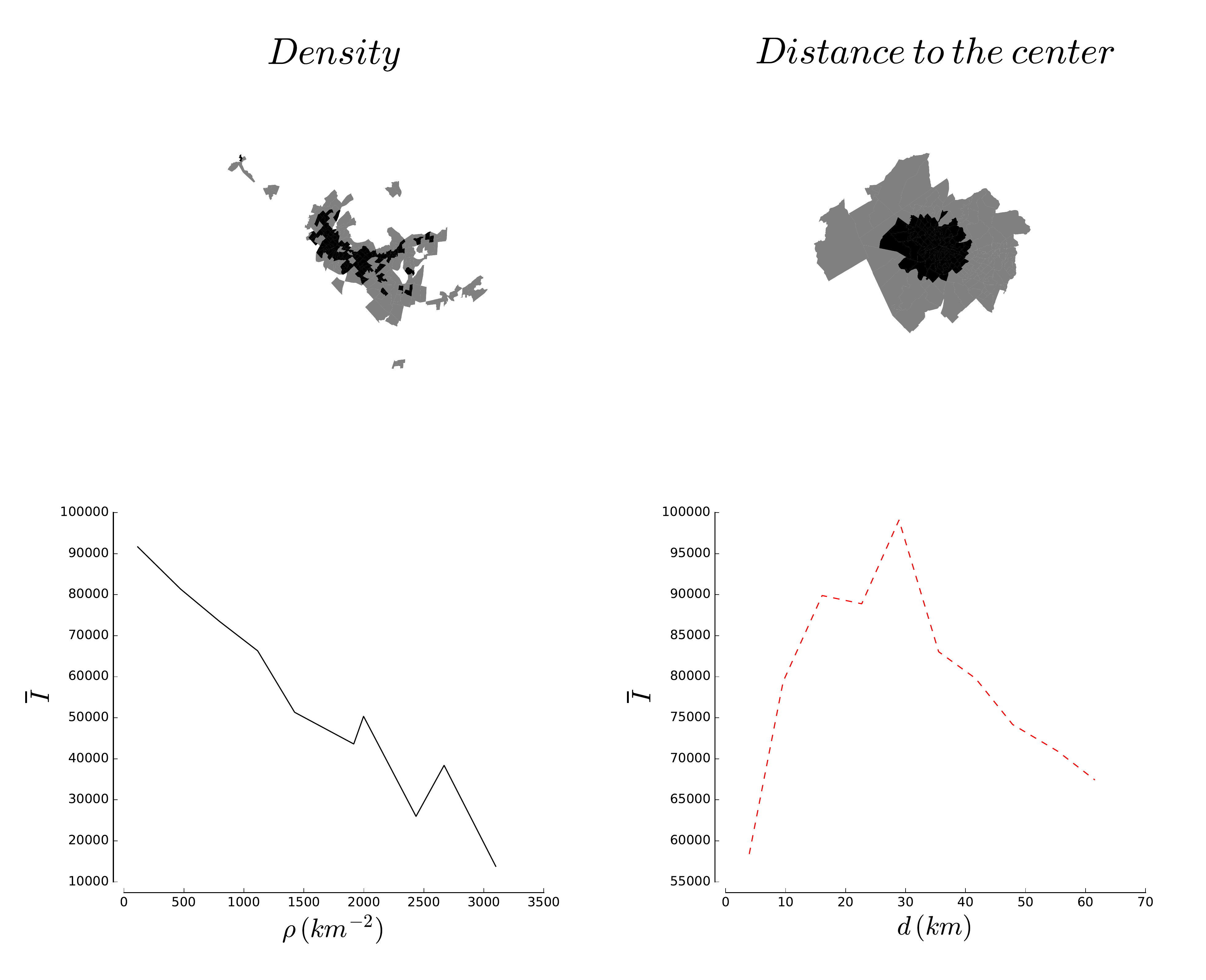}
\caption{{\bf Illustrating the density approach.} (Top Left) Blockgroups in
    Seattle, WA; the colour of a block depends on the population
    density interval to which it belongs (from the $25\%$ most dense in black to the $25\%$ least dense in
    white). (Top Right) Blockgroups in Seattle, WA coloured by distance to the
    center defined as densest blockgroup.  The colour goes from black
    for the $25\%$ closest blocks to white for the $25\%$ closest ones.  (Bottom Left) The average income of households as a
    function of density.  (Bottom right) Average income of households as a
    function of distance to the center. Both methods give very different
    results.  
    \label{fig:illustrate_percolate}}
\end{figure*}

The center of a city is usually defined as the region which has the
highest population (or employment) density. We therefore propose the
density as a proxy to measure of how `central' an area is.  We thus
plot quantities computed over all areal units (blockgroups in this
dataset) that have a density population in a given interval
$[\rho,\rho+d\rho]$ where $\rho$ decreases from its maximum to its
minimum value. We illustrate this idea and compare its results to the
traditional `distance to the center' method on
Fig.~\ref{fig:illustrate_percolate}. With very anisotropic and
polycentric cities such as Los Angeles, the order in which the areal
units are considered is very different with both methods. As a result,
measurements as simple as the average income will yield very different
results. This is particularly striking with the example of Seattle, WA
shown on Fig.~\ref{fig:illustrate_percolate}. The average income as a
function of the distance to the center (areal unit with the highest
density) increases from the center to a peak at roughly $30$ km
(bottom left figure). On the other hand, the bottom left figure shows
that average income is, in fact, a simple decreasing function of
residential density.

Where does the discrepancy come from? As one can see on the maps on the top of
Fig~\ref{fig:illustrate_percolate}, the units considered in a given distance range
can be very different in terms of density. Because real cities are neither
monocentric or isotropic, units at a same distance from the center can in fact
be very different. This shows, if anything, the importance of expliciting what
one means by `central' before presenting measures. In the following, we express 
centrality in terms of residential density.

\subsubsection{Results}
\label{sec:Results}

\begin{figure}
    \centering
    \includegraphics[width=0.49\textwidth]{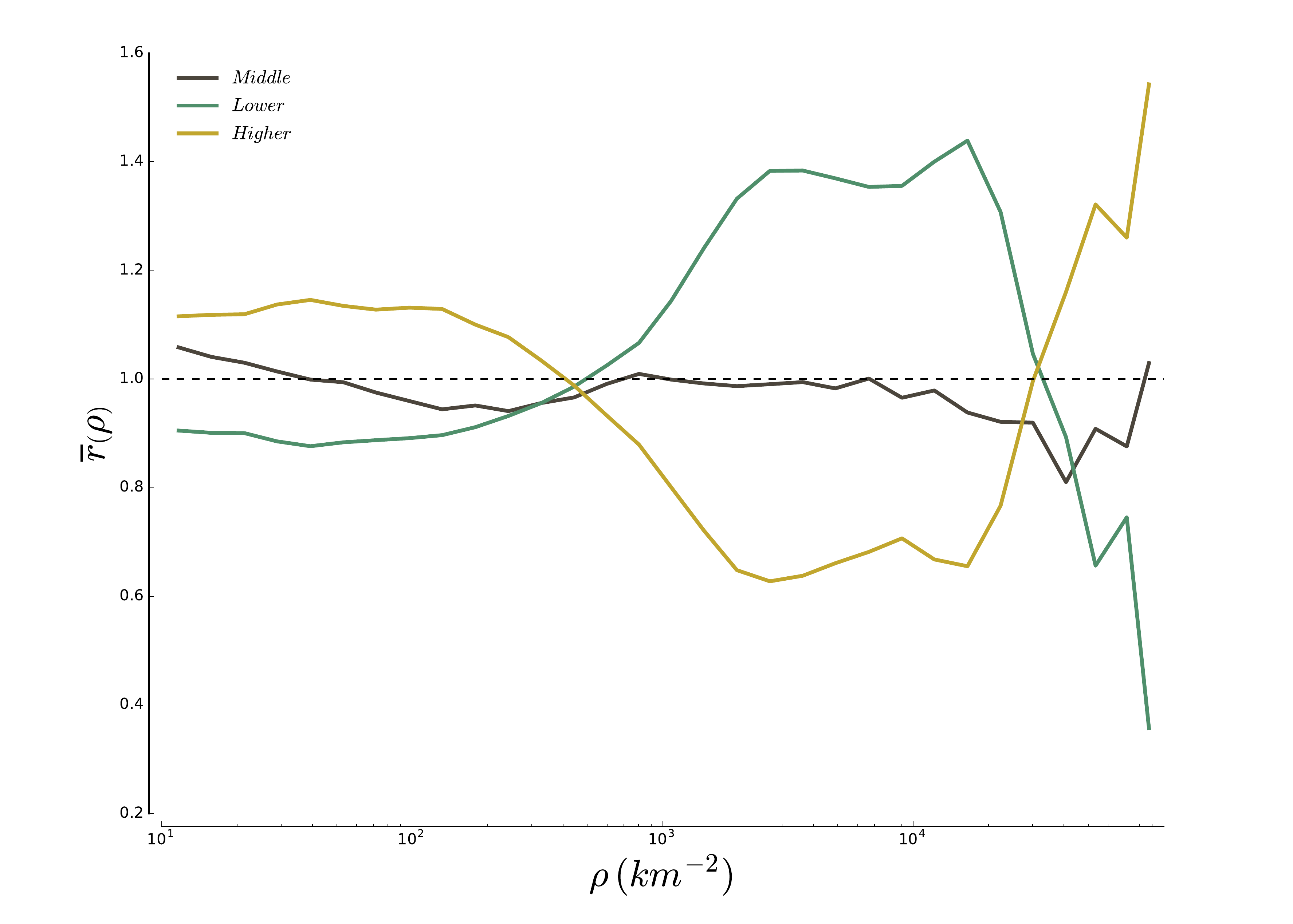}
    \caption{Average representation $\overline{r}$ of the higher-, middle- and
    lower-income classes over the $929$ CBSA as a function of the
    local density of households. On average, we find that low-density regions (the
    suburbs) are on the richer end, while high density regions (the center) are
    on the poorer end. This confirms on a large dataset a stylized fact that had previously
    emerged from local studies. Interestingly, we also
    find that high income households are on average overrepresented in very large density areas ($\rho>20,000/km^2$),
    suggesting that density may be one relevant factor in the explanation of the 
    differences between neighbourhoods.
    \label{fig:high_low_densities}} 
\end{figure}

Here, we compute the representation of groups as a function of
residential density. This method sheds a new light on the difference
of social composition between the high-density and low-density areas
in cities. Indeed, as shown on Fig.~\ref{fig:high_low_densities}, we
find that low-density regions in cities are on average rich
neighbourhoods and that higher density regions are on average
lower-income neighbourhoods, in agreement with the dichotomy rich
suburbs/poor centers usually found in the literature. But the
dichotomy is not the full picture. The method indeed entails a
surprising result: areas with very large densities (typically above
$10,000$ inhabitant$/\text{km}^2$) are on average rich
neighbourhoods. Only few cities in the US have neighbourhoods that
reach the threshold of $20,000$ inhabitants per km$^2$, which can
explain why people have reported in most cases the existence of poor
centers and rich suburbs. In fact, among all $630$ blockgroups with a
population density greater than $20,000$ inhabitants, $91\%$ are
located in New York, NY. Most high-density blockgroups belonging to
other metropolitan areas exhibit an overrepresentation of the
higher-income group and it is thus difficult to conclude at this stage
that we are observing an effect specific to Manhattan. In any case,
this result suggests that density is very relevant in the usual
discussion about income strucutre differences between north american
and european cities.

\subsection{Neighbourhoods and their properties}
    \label{sub:neighbourhoods_and_their_properties}

\begin{figure}
    \centering
    \includegraphics[width=0.49\textwidth]{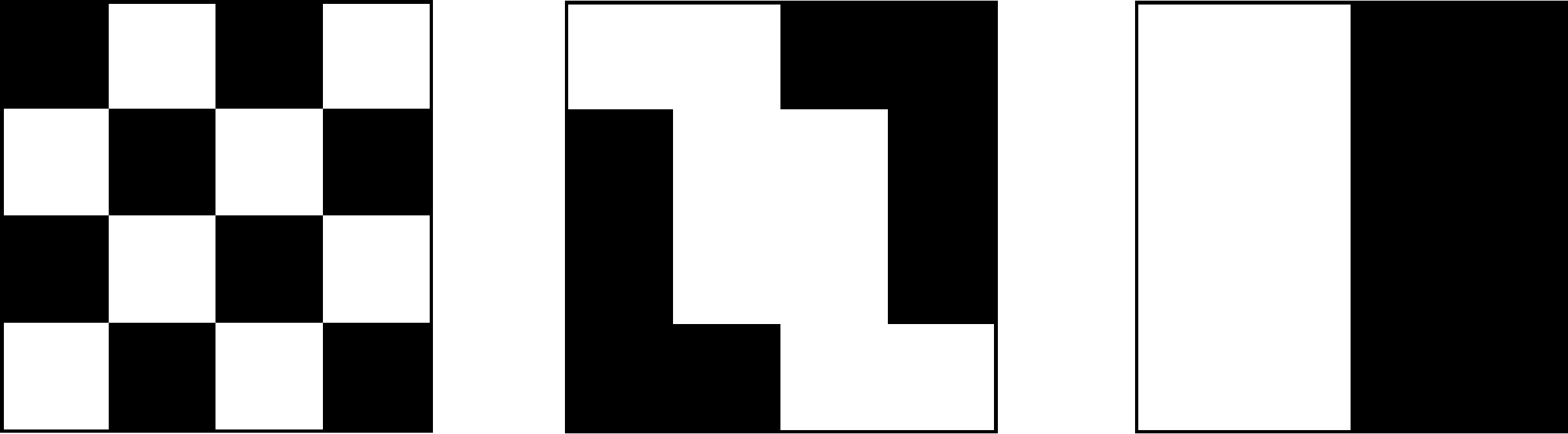}
    \caption{Three (hypothetical) spatial distributions of households that lead to the same
        values for intra-areal unit measures, but  represent different segregation patterns. (Left) The checkerboard
        city popularised by White~\cite{White:1983} corresponds to a
        clustering value---defined in Eq.~\ref{eq:clustering}---of $C=0$ for the black squares. (Middle) An
        intermediate situation between the checkerboard and the divided city,
        corresponding to $C \approx 0.86$. (Right) The divided city, corresponding to
        $C=1$. \label{fig:checkerboard}} 
\end{figure}

Intra-unit measures such as the representation or the exposure are not enough to quantify
segregation. Indeed, areal units where a given class is overrepresented can
arrange themselves in different ways, without affecting intra-unit measures of 
segregation~\cite{White:1983}. In order to illustrate this, we
consider the schematic cases represented on Fig.~\ref{fig:checkerboard}, and
assume that  they are obtained by reshuffling the various squares around.
Obviously, the checkerboard on the left depicts a very different segregation
situation from the divided situation on the right while intra-unit measures
would give identical results.

\subsubsection{Defining neighbourhoods}
\label{sec:Defining neighbourhoods}

Defining neighbourhoods in which categories tend to gather is a
difficult task.  Indeed, what individuals call neighbourhoods often
depend on their perception of the city, and field work is often
necessary to identify which areas are socially recognised as being a
large, middle or low income neighbourhood. However, it is often not
possible to do field work and finding a way to define neighbourhoods
from population counts is by more convenient and reliable.

What is usually defined as a neighbourhood defies the most naive
measures. For instance, to be a member of an `$\alpha$ neighbourhood'
(where $\alpha$ is here higher, middle or low income class) it is not
necessary for an area to have a majority of individuals from the class
$\alpha$~\cite{Logan:2011}. More sophisticated methods are thus
required and the literature is not exempt of such measures, that are
all rooted in different assumptions about the nature of
neighbourhoods~\cite{Shevky:1961, Sampson:1997, Logan:2011,
  Spielman:2013}. For instance, Logan et al.~\cite{Logan:2011} use
local K-functions in order to assess the prevalence of individuals
from a class in an area. The areas are then clustered using a standard
k-means clustering algorithm. The main issue with this approach is the
use of K-functions which measure absolute concentration and are based
on the null hypothesis of a completely random distribution of
individuals across space. As mentioned earlier in this manuscript, it
is more accurate to consider deviations from the null hypothesis of a
random distribution of individuals \emph{among existing locations}. We
thus propose here an improvement over Logan et al.'s definition based
on data given at the areal unit level (but could easily be generalised
to data with exact locations). As in~\cite{Logan:2011}, we start with
the intuitive idea that an $\alpha$ neighbourhood is an area of the
city where the category $\alpha$ is more present than in the rest of
the city. In other words, an areal unit $t$ belongs to an $\alpha$
neighbourhood if and only if the category $\alpha$ is overrepresented
in $t$, i.e. $r_\alpha(t) > 1$.  We then build neighbourhoods by
aggregating the adjacent areal units where the income class $\alpha$
is also overrepresented (see for example of Atlanta Fig.~\ref{fig:atlanta_neighbourhoods}).

\subsubsection{Clustering}
\label{ssub:clutering}

\begin{figure*}
    \centering
    \includegraphics[width=0.99\textwidth]{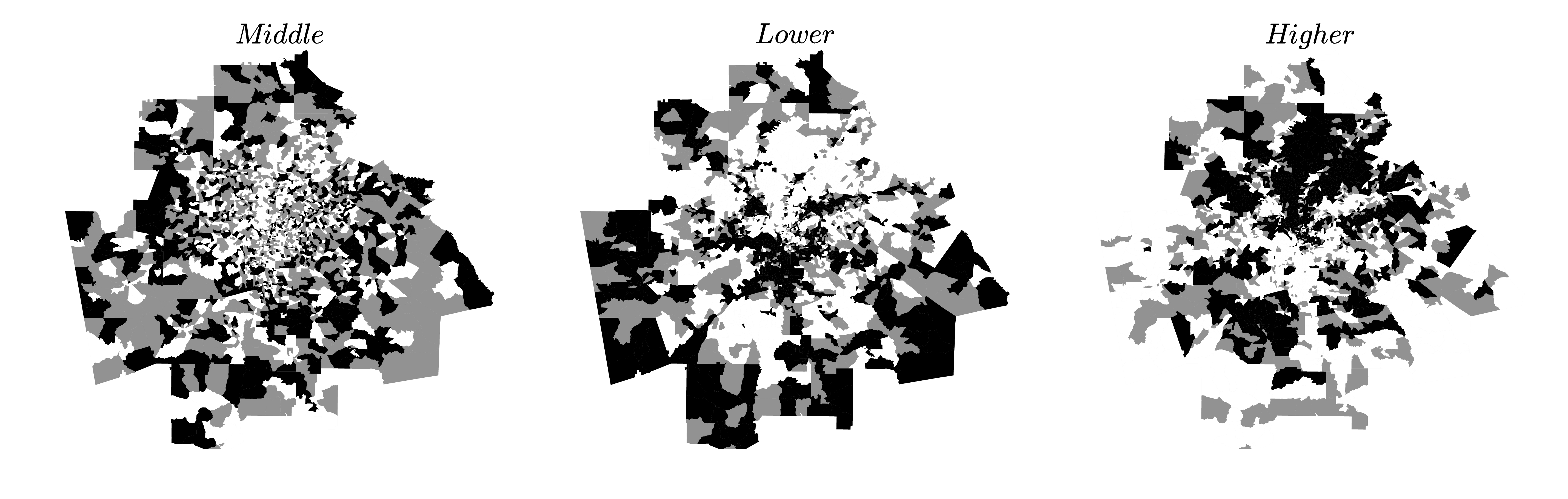}
    \caption{The neighbourhoods in Atlanta for the three income
      classes. In black, the blockgroups where the corresponding class
      is overrepresented; in white, where it is underrepresented; in
      grey, where the value of the representation is not
      distinguishable from the value that would be obtained if
      households chose their residence at random.  It is interesting
      to note that all CBSA defined for the $2014$ American Community
      Survey exhibit a total exclusion between lower-income and
      higher-income neighbourhoods: the pictures for lower- and
      higher-income classes are the perfect negative of one
      another. In contrast, middle-income households are scattered
      across the city and exhibit very little geographical coherence.}
\label{fig:atlanta_neighbourhoods}
\end{figure*}

\begin{figure} 
    \centering
    \includegraphics[width=0.49\textwidth]{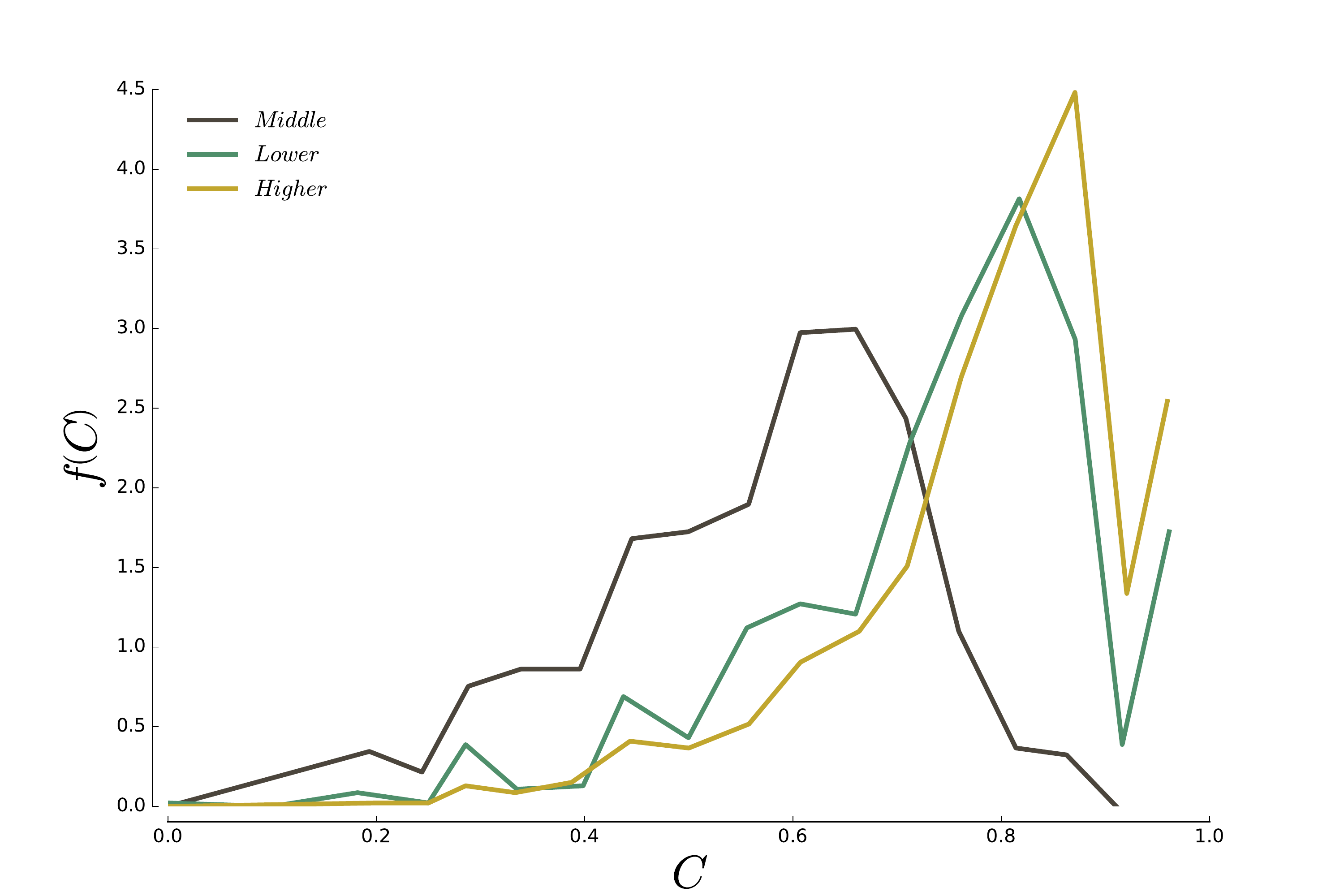}\\
    \caption{Distribution $f(C)$ of the value of the clustering
      coefficient $C$ per class for all cities in our dataset. The
      higher-income class exhibits the highest level of clustering,
      with an average of $\overline{C} = 0.83$, followed by the
      lower-income class with on average $\overline{C} = 0.78$. The
      middle-income class households are significantly less clustered
      than the other two, with $\overline{C} = 0.52$ on average. The
      average is computed over all US core-based statistical areas.}
        \label{fig:clustering} 
\end{figure}

A way to distinguish between different spatial arrangements is to
measure how clustered the overrepresented areal units are.  The ratio
of the number $N_n(\alpha)$ of $\alpha$-neighbourhoods (clusters) to
the total number $N_o(\alpha)$ of areal units when the class $\alpha$
is overrepresented (before constructing the neighbourhood as defined
above) gives a measure of the level of clustering and the quantity
\begin{equation} 
    C_\alpha = \frac{N_{o}(\alpha)-N_{n}(\alpha)}{N_{o}(\alpha)-1} 
    \label{eq:clustering}
\end{equation}
is such that $C_\alpha = 0$ in a checkerboard-like situation, and $C_\alpha = 1$
when all areal units form a unique neighbourhood. We show on
Fig.~\ref{fig:clustering} the distribution of $C_\alpha$ for the three classes over all
cities in our dataset. As one could infer from the maps on
Fig.~\ref{fig:atlanta_neighbourhoods}, the higher-income and lower-income areal units are well
clustered, with a respective average clustering of $C = 0.83$ and $C = 0.78$.
The middle class is on the other hand less spatially coherent, with a average clustering
$C = 0.52$.

\subsubsection{Concentration}
\label{ssub:concentration}

\begin{figure} 
    \centering
    \includegraphics[width=0.49\textwidth]{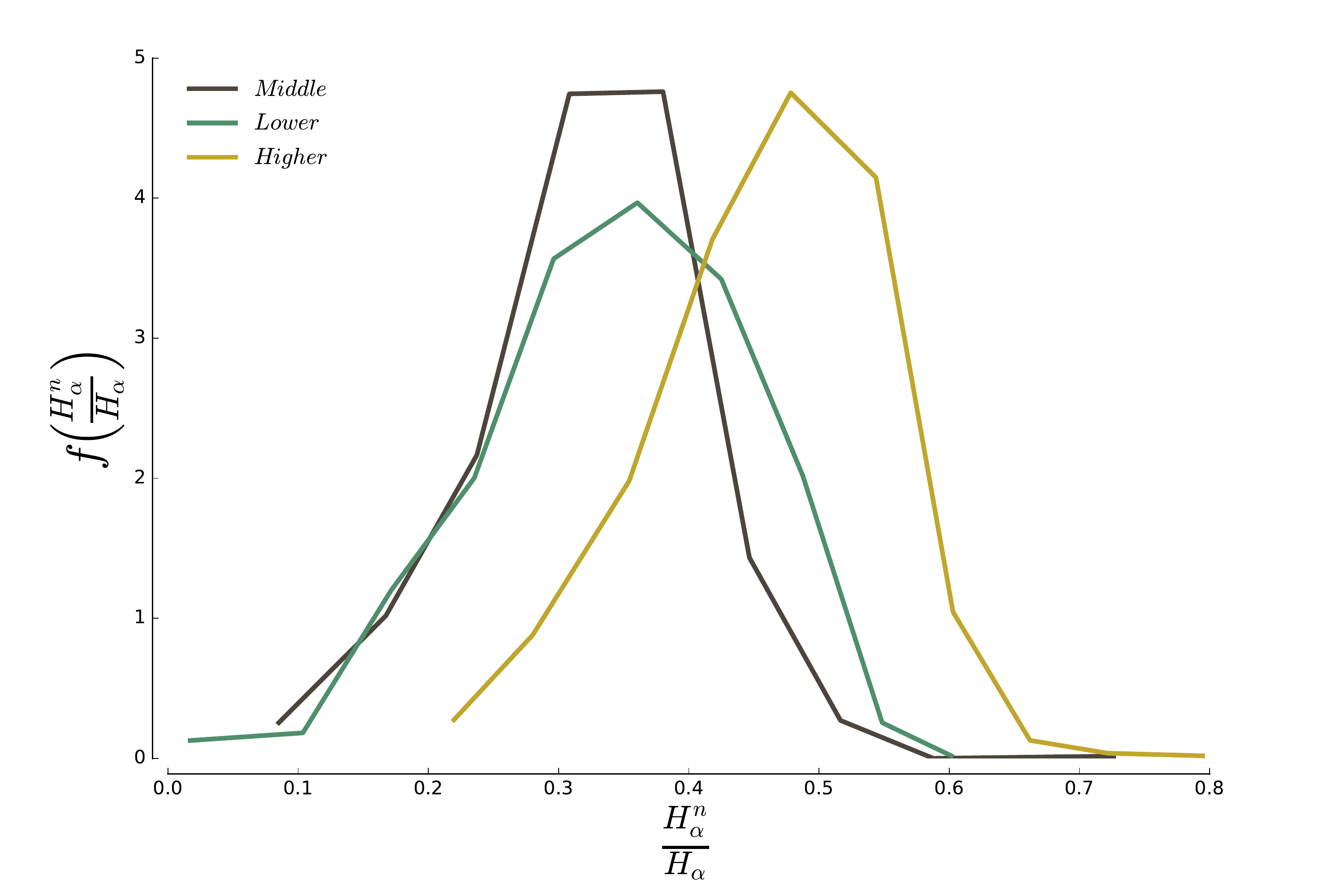}\\
    \caption{{\bf Concentration in neighbourhoods.} For each class $\alpha$ we compute
    the fraction of households $H_\alpha^n/H_\alpha$ that belongs to an $\alpha$ neighbourhood. The
    figure shows the distribution of this fraction for all 2014 CBSAs.} \label{fig:content} 
\end{figure}

If a given class is overrepresented in a neighbourhood, it does not
necessarily mean that most of the individuals belonging to this
neighbourhood are members of this class~\cite{Logan:2011}. Conversely,
the majority of individuals belonging to a class do not necessarily
all live in the above-defined neighbourhoods.  We thus compute the
ratio of households of each income class that lives in a neighbourhood
over the total number of individuals for that income class (rich,
poor, and middle class).  Our results (Fig.~\ref{fig:content})
indicate that about $50\%$ of the households belonging to $\alpha$ live in a
$\alpha$-neighbourhood, while the rest is dispatched across the rest
of the city. The average concentration decreases from higher-income
households ($52\%$), to lower-income ($40\%$) and middle-income
household ($40\%$).

\subsubsection{Fragmentation}
\label{ssub:fragmentation}

\begin{figure} 
    \includegraphics[width=0.49\textwidth]{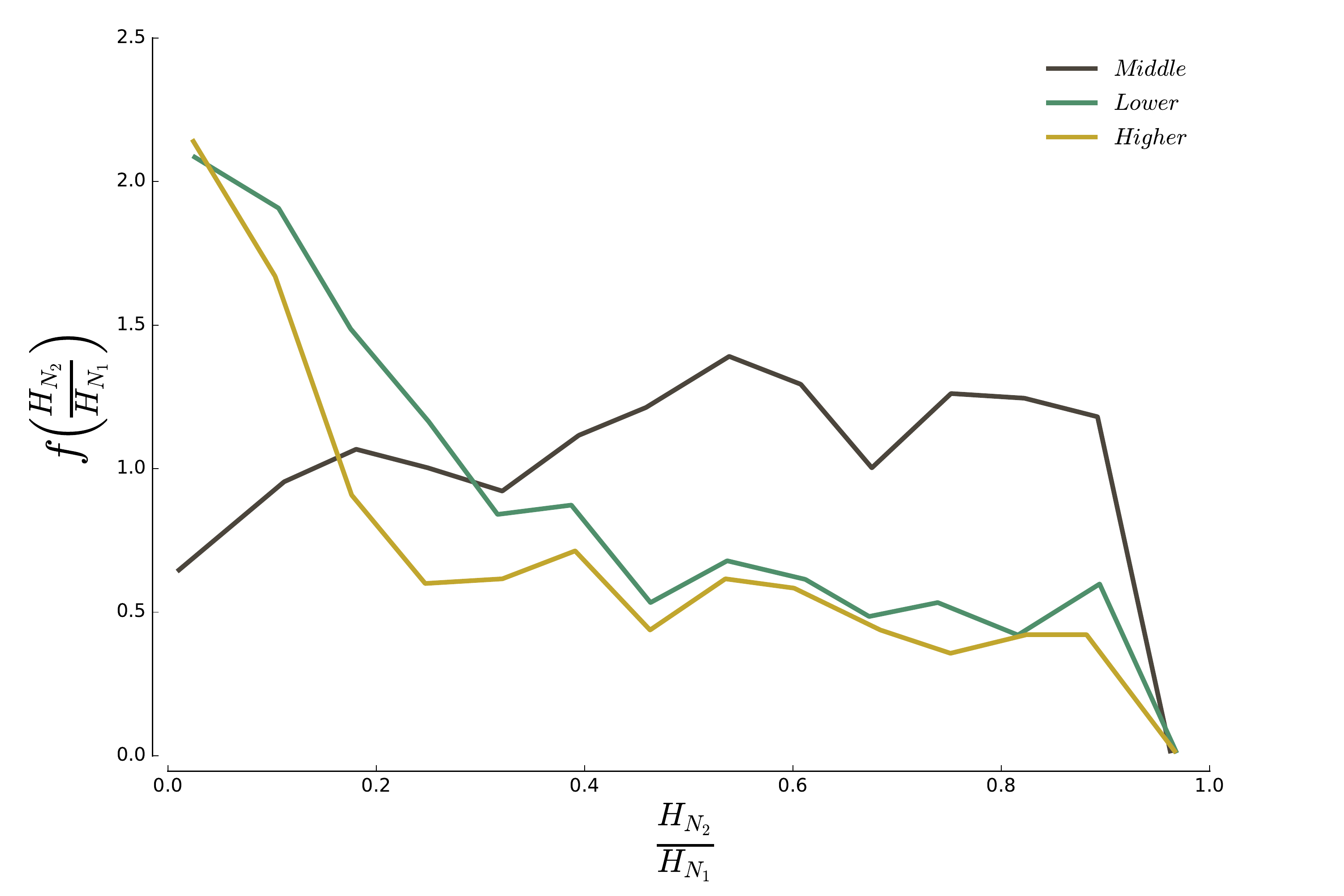}\\
    \caption{{\bf Neighbourhood fragmentation.} For each class $\alpha$ we
    compute the ratio of the size of the second largest
    $\alpha$-neighbourhood to the size of the largest $\alpha$-neighbourhood. The above figure
    shows the distribution of this ratio for all cities in our dataset.
    Higher- and lower-income households tend to concentrate in a single
    neighbourhood, with a secondary center that is respectively $27\%$ and
    $33\%$ the size of the largest one, on average over all cities. Middle-income
    households tend to be more dispersed, with a secondary neighbourhood that is on
    average $54\%$ of the size of the largest.} 
    \label{fig:polycentrism} 
\end{figure}

Finally, large values of clustering can hide different situations. We
could have on one hand a `giant' neighbourhood and several isolated areal units, which
would essentially mean that each class concentrates in a unique
neighbourhood. On the other hand, we could observe several neighbourhoods of
similar sizes, meaning that the different classes concentrate in
several neighbourhoods across the city. In order to distinguish
between the two situations, we plot
\begin{equation} 
    P = H_{N_2} / H_{N_1} 
\end{equation}
where $H_{N_1}$ is the population of the largest neighbourhood, and $H_{N_2}$
the population of the second largest neighbourhood. The results are shown on
Fig.~\ref{fig:polycentrism}, and again display a different behaviour for the
middle-income on one side, and higher-income and lower-income on the other side.
The size of the middle-income neighbourhoods are more balanced, with on average
$P=0.54$.  In contrast, higher- and lower-income neighbourhoods are
dominated by a single large neighbourhood with on average $P=0.27$ and $P=0.33$,
respectively.

\subsection{Larger cities are more segregated}
\label{ssub:dependence_on_city_size}
       
\begin{figure}
    \centering
    \includegraphics[width=0.49\textwidth]{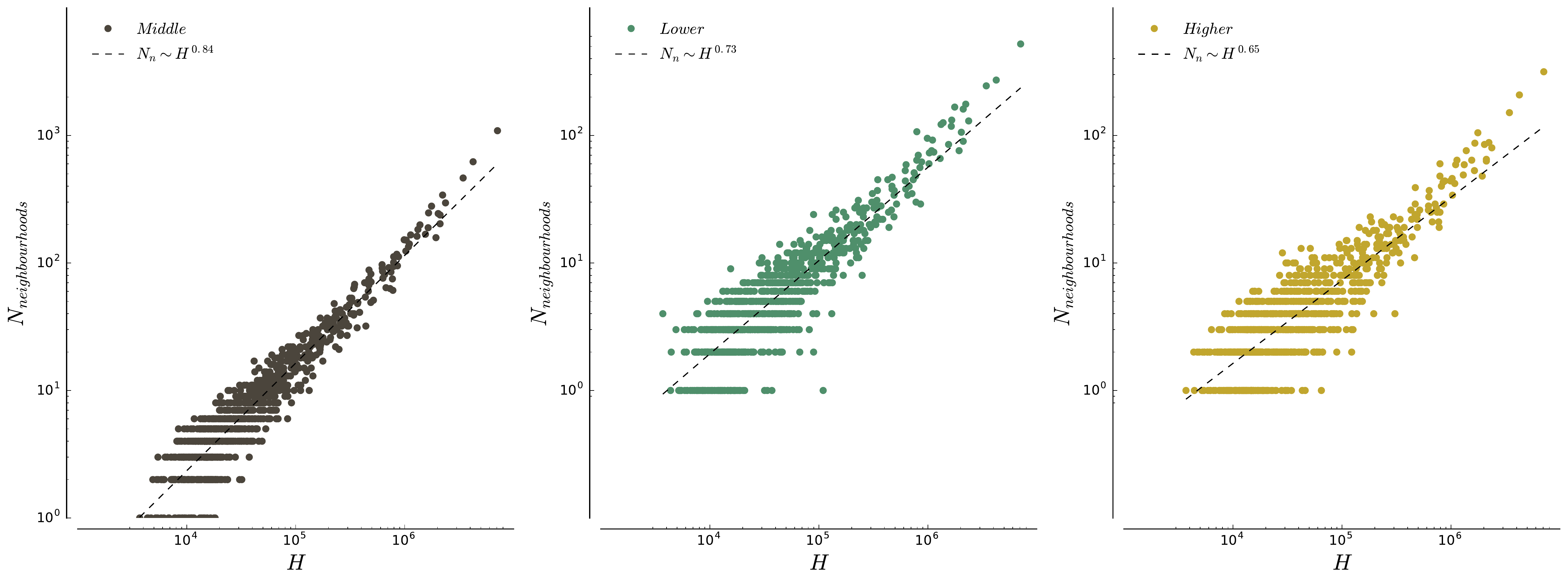}
    \caption{Number of neighbourhoods for the three different classes
      as a function of the size of the city. These plots in log-log
      show that we have a behavior consistent with a power law with
      exponent less than one (and with different value for each
      class). Combined with the linear increase of the number of
      over-represented units with the number of households (see
      Supplementary Information), this sublinear increase in the
      number of neighbourhoods shows the tendency of classes to
      cluster more as cities get
      larger.\label{fig:number_clusters_class}}
\end{figure}

As seen in Fig.~\ref{fig:clustering}, the clustering values are high, indicating
that the neighbourhoods occupied by households of different classes are sound.
We can now wonder whether there is an effect of the city size on the number of
neighbourhoods. We plot on Fig.~\ref{fig:number_clusters_class} the number of
neighbourhoods found for all three classes as a function of population. For each
class, The curve is well-fitted by a powerlaw function of the form

\begin{equation}
    N_n = b\,H^\tau
\end{equation}
where the exponent $\tau$ is less than one and depends on the class, 
indicating that there are proportionally less neighbourhoods
in larger cities (the number areal units scales proportionally with the
population size). In other words, different classes become more spatially
coherent as the population increases (see Supplementary Information below for more
details). The values of the exponents are
\begin{align*}
    \tau_{H} = 0.65\\
    \tau_{L} = 0.73\\
    \tau_{M} = 0.84\\
\end{align*}
These values show that the tendency to cluster as the city size increases is larger
for higher-income households than for lower- and middle-income households. In
other words, segregation increases with city size.

\section{Discussion}
\label{sec:discussion}

In this paper, we propose a general conceptual framework in which residential
segregation can be quantified and understood. Instead of  enumerating its
different aspects, we took a radically different -- yet simpler -- approach. We
define segregation by what it is not: a random distribution of the different
households throughout the urban space. This naturally leads to define the
measure of representation, which is used in turn to improve upon previous
ways~\cite{Logan:2011} to define neighbourhoods. We
further define the exposure (still based on the representation), which measures
the extent to which different categories attract, repel or are indifferent to
one another.

We show that we can define classes in a non-parametric way and 3 main
income classes emerge for the 2014 American Community Survey data. The
middle-income class corresponds to a smaller income range than what is
usually admitted, a curiosity that certainly deserves further
investigations. These complex systems can thus be described by
considering a small number of categories only. This is an important
piece of information which will simplify the description and modelling
of stratification mechanisms.

In terms of spatial arrangement, although the fraction of the population that is
contained in neighbourhoods does not change with city size, the neighbourhoods
are geographically more coherent as cities get larger, which corresponds in
effect to an increased level of segregation as the city size increases. 

Our results also point to the intriguing fact that higher-income households are on
average overrepresented in very dense areas. Such high density areas are
relatively rare in the US, which might explain in part why we observe poor
centers and rich suburbs and rich centers essentially in Europe where the
density is very large. This result echoes Jacobs' analysis~\cite{Jacobs:1961}
that neighbourhoods with the highest dwelling densities are
usually the ones exhibiting the largest vitality, and are
therefore the most attractive. Obviously, a high density is not the only
determinant and in some cases high-density neighbourhoods can also be 
lower-income neighbourhoods. Further investigations along these lines may
provide quantitative insights into the mechanisms leading to urban decline or
urban regeneration. 

In this study, we also have tried to highlight the \emph{spatial} pattern of
segregation. We believe that the identification of neighbourhoods that our
method permits will allow a finer-scale investigation of these spatial patterns.
However, the fundamental issue that runs beneath -- the need for new tools in the
analysis of spatial patterns -- is still open. It goes beyond the problem of
segregation and has a huge number of potential applications.

\section{Materials and methods}
\subsection{Data}
In this study, we use the American Census Bureau's 2014 American
Community Survey data on the income of households at the census
block-group level, grouped per Core-based Statistical Areas. The
households are divided in 16 income categories, ranging from below
$\$10,000$ annual income to above $\$200,000$.  All data of the 2014
American Community Survey are available from the Census Bureau. 2140
delineations of the Core-based Statistical Areas are available from
the Office of Budget Management. The reader interested in obtaining a
cleaned version of these data ready for analysis and/or reproduce the
results of this analysis can consult the online repository (The code
necessary to download, assemble the data and reproduce the analysis
performed in this article is freely available online at
http://github.com/rlouf/patterns-of-segregation).

\subsection{Software}
The methods described in this manuscript are very general, and not limited to
the study of income segregation. In order to facilitate their application to
other datasets, all the measures have been packaged in a python library, Marble,
open-source and freely available online~\cite{github:marble}.

\section{Supplementary Materials}

The following supporting information provide more details on the calculations
made to obtain the maximum and minimum value of exposure and isolation, and
their variance. We also detail the process through which we agregate the
original categories into classes, and the results we obtain for the $2014$
American Community Survey. Finally, we discuss in more details some of the
results presented in the manuscript; in particular, our claim that larger cities
are richer.

\section{Exposure}

    \subsection{Definition}

Given two different categories $\alpha$ and $\beta$, we define their
intra-unit exposure as the average representation $r_{\alpha}$ (resp.
$r_\beta$) of the $\alpha$ (resp. $\beta$) that is seen on average by the
members of $\beta$ (resp. $\alpha$)

\begin{equation}
	E_{\alpha \beta} = \frac{1}{N_\alpha} \sum_t  n_{\alpha}(t)\: r_\beta(t)	
\end{equation}

It can also be re-written

\begin{equation}
	E_{\alpha \beta} = \frac{1}{N} \sum_{t=0}^{T} n(t)\, E_{\alpha \beta}(t)
\end{equation}

where

\begin{equation}
	E_{\alpha \beta}(t) = r_\alpha(t)\,r_\beta(t)
\end{equation}

This expression symmetric by permutation of the categories $\alpha$ and $\beta$, which is
what one would expect from an index measuring the interaction between two
categories.

    \subsection{Expected value and variance in the random configuration}

In order to know whether the attraction or repulsion mesured between two classes
is significant, we need to be able to compute $\Var\left[E_{\alpha
\beta}\right]$. Assuming that $r_{\alpha}(t)$ and $r_{\beta}(t)$ are independent
(which is rigorously not true for tracts with a fixed capacity $n(t)$), it
follows

\begin{widetext}
\begin{align*}
	\E\left[ E_{\alpha \beta}(t)\right] &= \E[r_{\alpha}(t)] \E[r_{\beta}(t)]=1\\
	\Var[ E_{\alpha \beta}(t)] &= \frac{1}{N_\alpha N_\beta} \left(
\frac{N}{n(t)}-1 \right)^2 + \frac{1}{N_\alpha} \left( \frac{N}{n(t)} - 1
\right) + \frac{1}{N_\beta} \left( \frac{N}{n(t)} - 1
 \right)
\end{align*}

Thus 

\begin{align*}
	\E[E_{\alpha \beta}] &= 1\\
	\Var[ E_{\alpha \beta} ] &= \frac{1}{N^2} \sum_t n(t)^2\, \Var[E_{\alpha
\beta}(t)] + \frac{2}{N^2} \sum_{s<t} n(s)\, n(t)\, \Cov[ E_{\alpha \beta}(s), E_{\alpha
\beta}(t)]
\end{align*}

The covariance is non-zero because the $n_{\alpha}(t)$ of two different tracts
$t$ and $s$  are not independent, and we have

\begin{equation}
	\Cov[E_{\alpha \beta}(s), E_{\alpha \beta}(t)] =
 \left(1-\frac{1}{N_\alpha}\right) \left(1-\frac{1}{N_\beta}\right) - 1
\end{equation}
\end{widetext}

    \subsection{Minimum and maximum values}

In order to be able to make sense of the values of exposure ($E_{\alpha \beta}$) and isolation
($I_{\alpha}$),  and compare different cities, we need to know their respective maximum
and minimum values. We will consider the following cases:

\begin{description}
    \item[Maximum isolation] 
            Situation where each areal unit contains households from one and only
            one category. This situation corresponds to the minimum of $E_{\alpha \beta}$ 
            and the maximum of $I_{\alpha}$.
    \item[The unsegregated city] 
            When the distribution of households in the different areal units
            cannot be distinguished from a random distribution. This
            is what we call the `unsegregated city' and gives a point
            of reference. It corresponds to the minimum of $I_{\alpha}$.
\end{description}

\subsubsection{Isolation $I_\alpha$}

In the unsegregated city case, there is no way to tell
the difference between the distribution of the different categories in the
different tracts and a random distribution. In this situation, isolation indices
$I_{\alpha}$ reach their minimum value
\begin{equation*}
    \boxed{I^{\,min}_{\alpha} = 1} 
\end{equation*}
when $r_\alpha(t) = 1,\; \forall\, t$.

In the maximum isolation case, all categories are alone in their own tract. In other words,
$\forall\,t$ and $\forall\, \beta\neq\alpha$ we have $n_\beta(t)=0$ iff
$n_\alpha(t)\neq0$. We thus obtain for the isolation
\begin{align*}
	I_{\alpha} &= \frac{1}{N_\alpha} \sum_{t=1}^{T}
	n_\alpha(t)\,r_\alpha(t)\\
        &= \frac{1}{N_\alpha} \sum_{t \in \mathcal{R}_\alpha}
        \frac{n_\alpha(t)^2}{n(t)}
	\frac{1}{N_\alpha / N} \\
        &= \frac{N}{N_\alpha^2} \sum_{t \in \mathcal{R}_\alpha}
        \frac{n_\alpha(t)^2}{n(t)}
\end{align*}
where $\mathcal{R}_\alpha$ is the set of areal units where the category $\alpha$
is present. In these unit, $n(t) = n_\alpha(t)$. Therefore
\begin{equation*}
    \boxed{I^{\,max}_{\alpha} = \frac{N}{N_\alpha}} 
\end{equation*}

\subsubsection{Exposure $E_{\alpha \beta}$}

In the maximum isolation case, all categories are alone in their own tract. In other words,
$\forall\,t$ and $\forall\, \beta\neq\alpha$ we have $n_\beta(t)=0$ iff
$n_\alpha(t)\neq0$. In this situation, we trivially have

\begin{equation*}
    \boxed{E^{\,min}_{\alpha \beta} = 0}
\end{equation*}

The maximum of the exposure is however more difficult to obtain in
general. We fix $\alpha$ and $\beta$ and we denote by a category $\gamma$ all the
rest. By definition we have $\sum_tn_\alpha(t)=N_\alpha$,
$\sum_tn_\beta(t)=N_\beta$, and
$\sum_tn_\gamma(t)=N-N_\alpha-N_\beta$. 

We will look for the `global' maximum by keeping the only constraint that in each unit we  have
$n(t)=n_\alpha(t)+n_\beta(t)+n_\gamma(t)$. We obtain for the exposure
\begin{equation}
E_{\alpha\beta}=\frac{N}{N_\alpha
  N_\beta}\sum_t\frac{n_\alpha(t)(n(t)-n_\alpha(t)-n_\gamma(t))}{n(t)}
\end{equation}
The maximization of the exposure with respect to $n_\alpha(t)$ thus gives
\begin{equation}
\frac{\partial E_{\alpha\beta}}{\partial
  n_\alpha(t)}=0=\frac{N}{N_\alpha
  N_\beta}\left[n(t)-n_\gamma(t)-2n_\alpha(t)\right]
\end{equation}
which leads to
\begin{equation}
n_\alpha^*(t)=\frac{n(t)-n_\gamma(t)}{2}
\end{equation}
The exposure for these values reads
\begin{equation}
E_{\alpha\beta}(\{n^*_\alpha\},\{n_\gamma\})=\frac{N}{N_\alpha
  N_\beta}\sum_t\frac{(n(t)-n_\gamma(t))^2}{4n(t)}
\label{eq:expomax}
\end{equation}
The quantity $n_\gamma(t)$ is in the compact set $[0,n(t)]$ and 
the maximization is not necessarily given by taking the
derivative equal to zero. Indeed, in this case the maximum of
$E_{\alpha\beta}(\{n_\gamma\})$ is obtained for $n_\gamma(t)=0$ for
all $t$ (while the derivative equal to zero would lead to the minimum
obtained for $n_\gamma =n(t)$ for all t) and reads
\begin{equation}
    \boxed{E^{\,max}_{\alpha \beta} = \frac{N^2}{4N_\alpha N_\beta}} 
\end{equation}
This maximum is the global one, obtained when there are no
constraints. One can easily add the constraint
$\sum_tn_\alpha(t)=N_\alpha$ by using a Lagrange multiplier $\lambda$ and we
have then to maximize the function

\begin{widetext}
\begin{equation}
E_{\alpha\beta}=\frac{N}{N_\alpha
  N_\beta}\sum_t\frac{n_\alpha(t)(n(t)-n_\alpha(t)-n_\gamma(t))}{n(t)}-\lambda(\sum_tn_\alpha(t)-N_\alpha)
\end{equation}
\end{widetext}

The derivative with respect to $n_\alpha(t)$ leads to 
\begin{equation}
n_\alpha^*(t)=\frac{1}{2}\left(n(1-\frac{N_\beta-N_\alpha}{N})-n_\gamma\right)
\end{equation}
where we expressed the constraint $\sum_tn_\alpha(t)=N_\alpha$ in order
to eliminate the Lagrange multiplier $\lambda$. We can then express
the maximum $E_{\alpha\beta}(\{n_\gamma\})$ obtained for these values
of $n_\alpha$ and as above the maximum is obtained for $n_\gamma(t)=0$
for all t and reads
\begin{equation}
    E^{max,c}_{\alpha \beta} = \frac{N^2}{4N_\alpha N_\beta}\left[1-\left(\frac{N_\beta-N_\alpha}{N}\right)^2\right]
\end{equation}
which is obviously smaller than the global maximum $E^{max}_{\alpha\beta}$.

These maxima were obtained when there are no constraints on the total
number $\sum_tn_\gamma(t)$. When there is such a constraint, the
construction of the maximum of Eq. (\ref{eq:expomax}) is not trivial. Very likely, when
$\sum_tn_\gamma(t)=N_\gamma$ is fixed, we have to fill the smallest
tracts with this class $\gamma$ and we are then left with the classes
$\alpha$ and $\beta$ only. It seems difficult to obtain an analytical derivation of this maximum
and we will keep as a reference in our calculations the global maximum $E^{max}_{\alpha\beta}$.

\section{Agregating categories into classes}

The study of segregation must be rooted in a particular definition of
class.  However, the income is a continuous variable, and there is no
clear definition of incomes classes in the litterature: a class means
different things to different people. We thus start by finding out the
class structure as it manifests itself in the spatial arrangement of
people.

\subsubsection{Method}

We take as a starting point the finest income subdivision given by the Census Bureau ($16$
subdivisions) and compute the $16 \times 16$ matrix of $E_{\alpha \beta}$ values
at the scale of each cities. We then perform hierarchical clustering on this matrix, successively aggregating
the subdivisions with the highest $E_{\alpha \beta}$ values. The process,
implemented in the library Marble~\cite{github:marble},  goes as follows:

\begin{enumerate}
    \item Check whether there exists a pair $\alpha$, $\beta$ such that
        $E_{\alpha \beta} > 1 + 10\,\sigma$ (i.e. two categories that attract
        one another with at least 99\% confidence according to the Chebyshev
        inequality). If not, stop the agregation and return the classes;
    \item If there are some couples satisfying (1), normalize all $E_{\alpha \beta}$ values by their
        respective maximum values. Find then the pair $\gamma$, $\beta$ whose
        normalized exposure is the maximum;
    \item Aggregate the two categories $\beta$ and $\gamma$;
    \item Restart the process until it stops.
\end{enumerate}

In order to aggregate the categories at step 3, we need to compute the distance
betweence  $\delta = \beta \cup \gamma$ and any category $\alpha$ once $\beta$ and
$\gamma$ have been aggregated. Using the definition of $E_{\alpha \beta}$, it is
easy to show that 

\begin{equation}
E_{\alpha \delta} = \frac{1}{N_\beta + N_\gamma} \left( N_\beta\, E_{\alpha \beta}
+ N_\gamma\, E_{\alpha \gamma} \right)
\end{equation}

The variance is also easily calculated as:

\begin{equation}
	\Var \left[ E_{\alpha \delta} \right] = \frac{1}{\left( N_\beta +
	N_\gamma \right)^2} \left( N_\beta^2\,\Var \left[ E_{\alpha \beta}
	\right] + N_\gamma^2\, \Var \left[ E_{\alpha \gamma} \right] \right)
\end{equation}


\subsubsection{Computing the class structure at the country scale} 

We computed the class structure at the scale of the whole US. We
assume that the country is a juxtaposition of the different cities, with
independent values of $E_{\alpha \beta}^c$. We then compute the
average over the whole country and obtain
\begin{equation}
	E_{\alpha \beta}^{US} = \frac{1}{N_{US}} \sum_{c}^{} N_c\, E_{\alpha \beta}^c
\end{equation}

where $N_c$ is the population of the city $c$, and $N_{US}$ the urban population
of the US. The sum runs over all MSAs in the US. The variance is then
given by
\begin{equation}
	\Var \left[ E_{\alpha \beta}^{US} \right] = \frac{1}{N_{US}^2}
	\sum_{c}^{}(N_c)^2\, \Var \left[ E_{\alpha \beta}^c \right]
\end{equation}

\subsubsection{Results}

Starting from categories $0$ (the poorest) to $15$ (the wealthiest), our methods
finds the following classes for the US

\begin{align*}
    \text{(L - 59\%) }&0|1|2|3|4|5|6|7|8\\
     \text{(M - 11\%) }&9|10\\
     \text{(H - 29\%) }&11|12|13|14|15
\end{align*}

with in parenthesis the percentage of the total US population that is included in
the corresponding classes.

\section{Larger cities are richer and more unequal}

\begin{figure}
    \centering
    \includegraphics[width=0.5\textwidth]{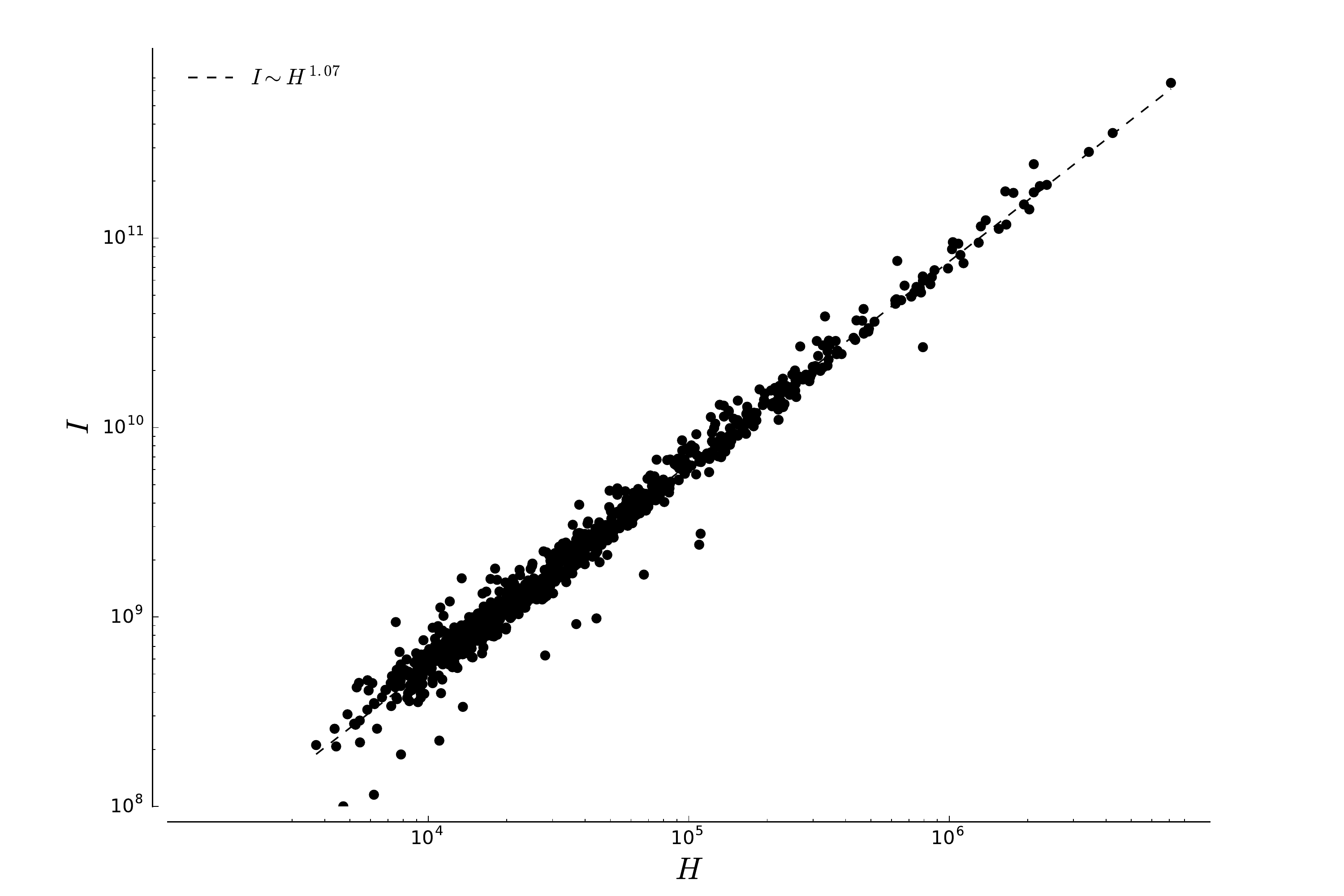}
    \caption{Total income of households versus the total number of households.
    The dashed line represents a power-law fit with exponent $1.08$ ($R^2 = 1$).
\label{fig:scaling_income}}
\end{figure}

Although intuitively appealing, the idea that larger Metropolitan areas are
richer is not as straightforward as it seems. The first question
one can  ask is if people are richer on average in large cities ? As
shown in Fig.~\ref{fig:scaling_income}, the total income in a city scales
(slightly) superlineary
with population size
\begin{equation}
    I \sim P^{1.07}
\end{equation}
which suggests that the income \emph{per household} is on average higher in larger
cities than in smaller ones. In other words, there are proportionally
more households belonging the to the wealthiest categories in large cities. In
other words, the income inequality is higher in large cities than
in small ones.

In order to measure levels of income inequality, we compute the Gini
coefficient of the income distribution for every Core-based Statistical Area
using the formula proposed in~\cite{Dixon:1987}
\begin{equation}
    G = \frac{1}{2\,N(N-1)\,\overline{I}} \sum_{i,j=1}^H \left| I_i - I_j
    \right|
\end{equation}

\begin{figure}[!h]
    \centering
    \includegraphics[width=0.5\textwidth]{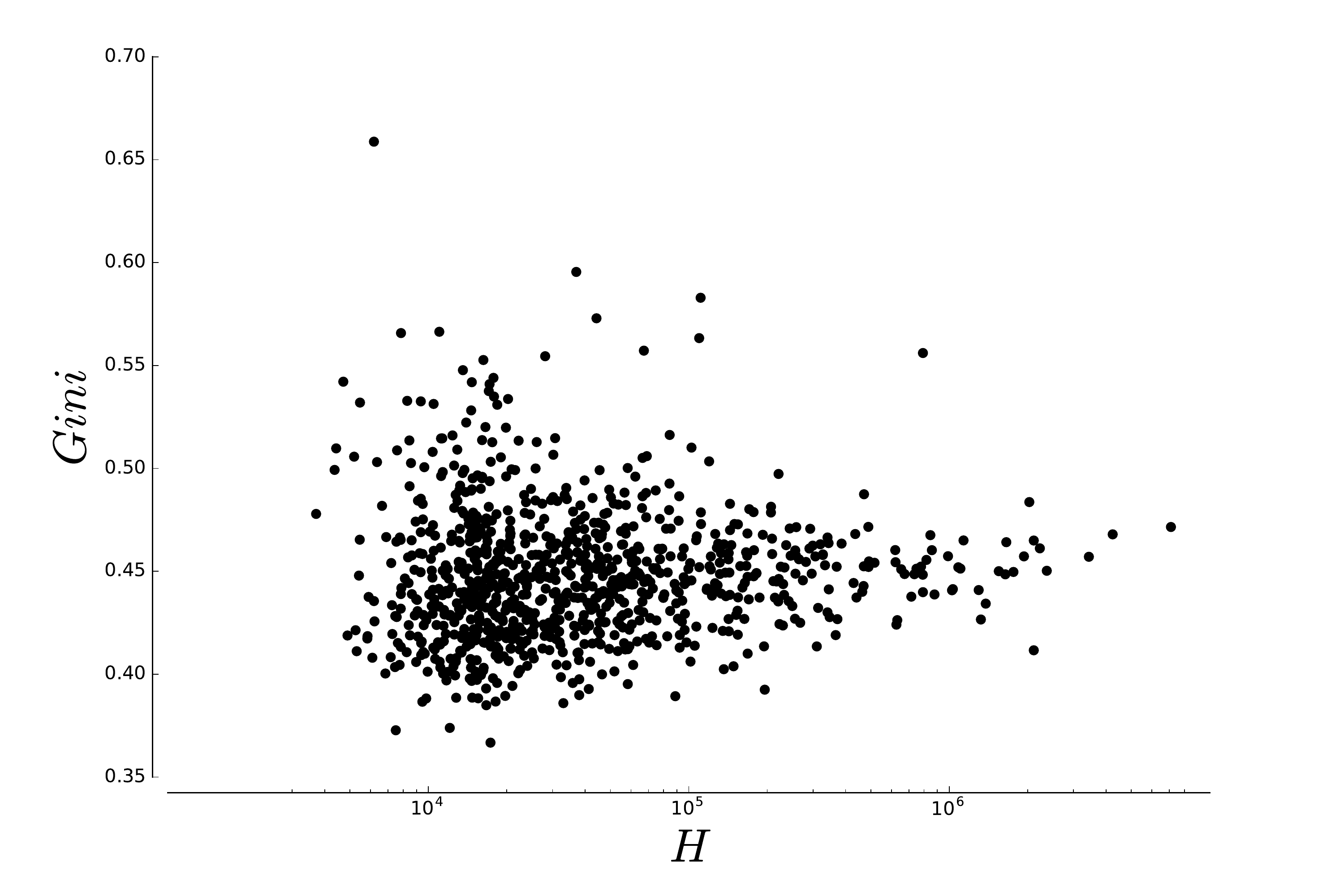}
    \caption{Gini coefficient of the income distribution of the $280$ MSA in
    $2014$ versus the number of households in the city. No clear trend
    can be observed here.\label{fig:gini}}
\end{figure}

The results, shown in Fig.~\ref{fig:gini} do not show any dependence
of the Gini coefficient on the metropolitan population. This example
shows that the Gini coefficient is not always a good measure of
inequality and can be too aggregated to detect finer details.
\begin{figure*}[!h]
    \centering
    \includegraphics[width=0.95\textwidth]{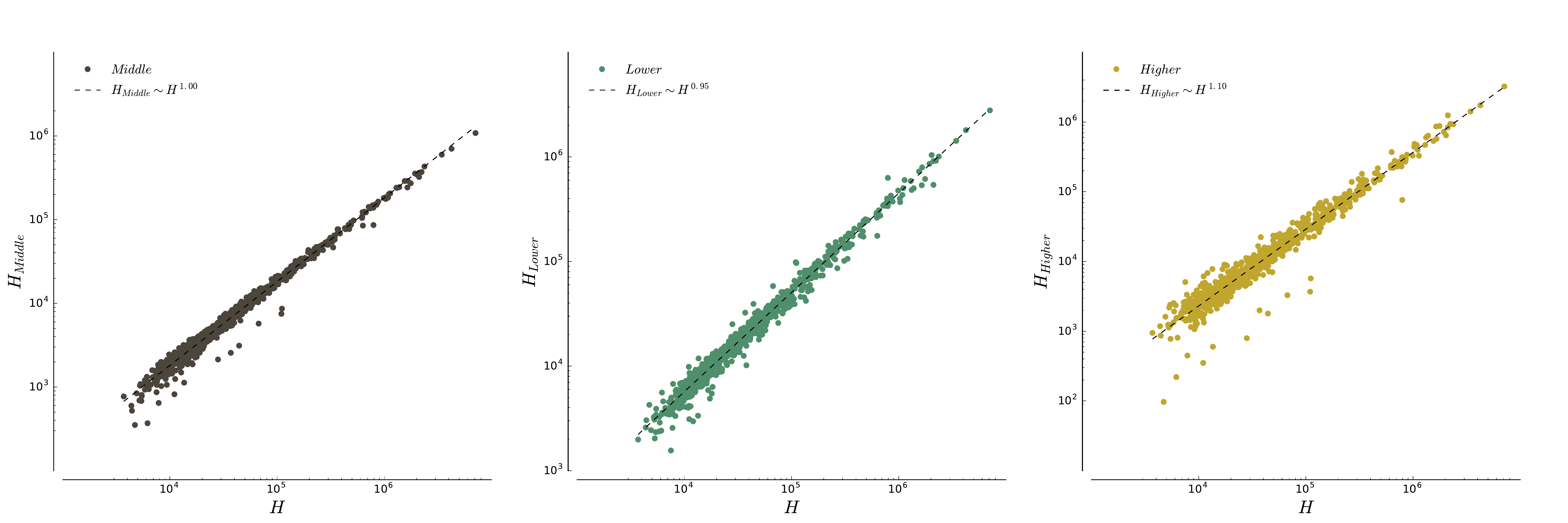}
    \caption{Scaling of the number of households in each class with the total
    number of households for the $2014$ Core-based Statistical Areas. The fits are extremely good with a $R^2 >
    0.98$ in all three cases.\label{fig:scaling_class}}
\end{figure*}
In order to confirm the consequence of the superlinear scaling of income in terms of
larger cities having proportionally more higher-income households, we plot the
number of households belonging to the $3$ different classes as a function of the
total number of households on Fig.~\ref{fig:scaling_class}. We find that for
three classes, the data are well approximated by a power-law
relationship
\begin{align*}
    H_{L} &\sim H^{\,0.95}\\
    H_{M} &\sim H^{\,1.00}\\
    H_{H} &\sim H^{\,1.10}\\
\end{align*}
The problem with writing scaling relationships in this case is that it
the constraint $H_L + H_M + H_H = H$ is hidden (ie. the numbers of households belonging
to each category must sum to the total number of households). We
therefore write 
\begin{equation}
    H_i = \eta_i(H)\, H\\
\end{equation}
where $\eta_i$ is the fraction of households in the city that belong to the
class $i$. The constraint that the numbers of households in each class should
sum to $H$ is equivalent to 
\begin{equation}
    \eta_L + \eta_M + \eta_H = 1
\end{equation}

\begin{figure*}
    \centering
    \includegraphics[width=0.95\textwidth]{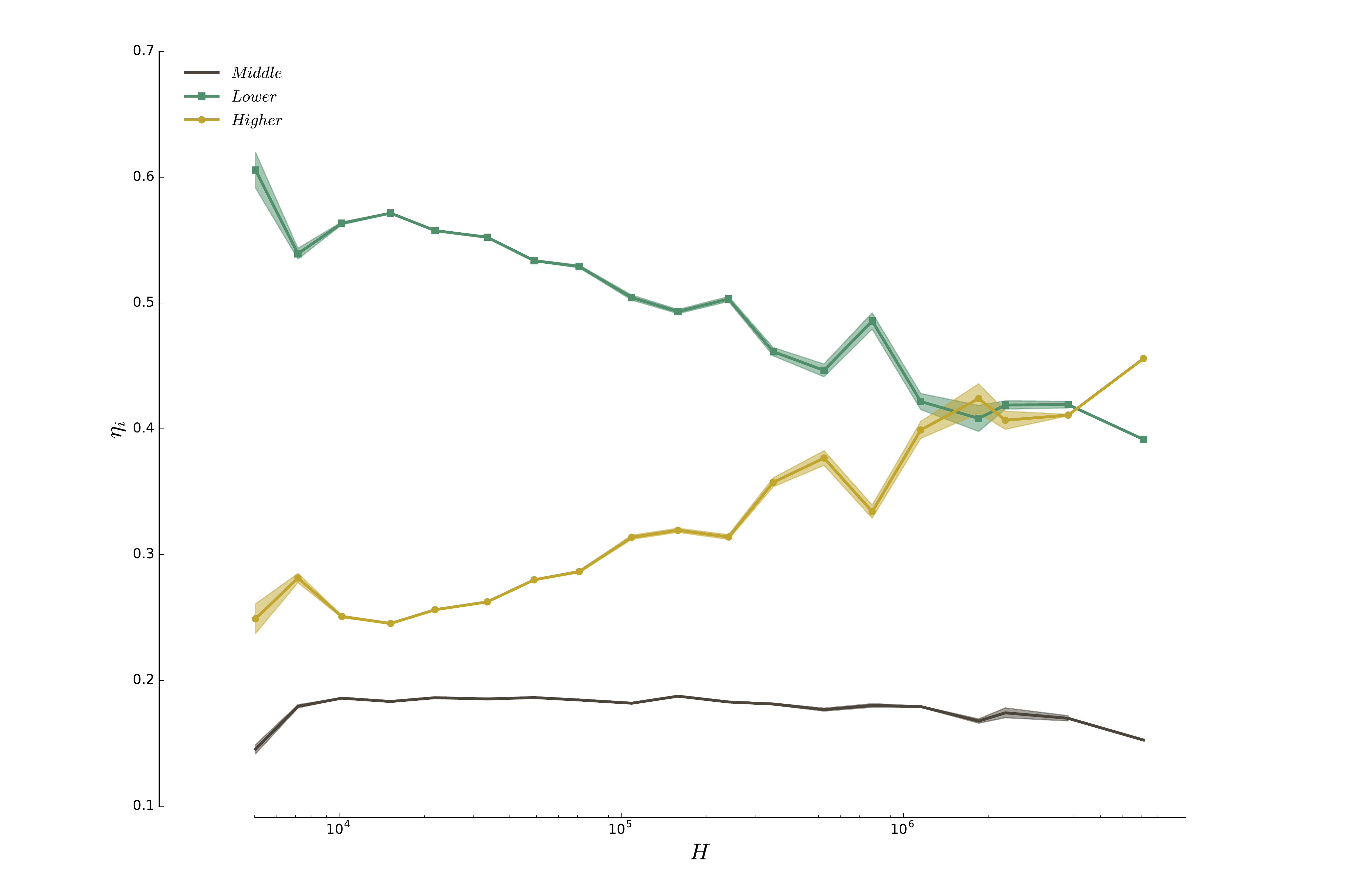}
    \caption{Fraction $\eta_i$ of the number of households belonging to the
    class $i$ in cities versus the total number $H$ of households. The curve shows the
average values obtained when binning per household (the shaded area represent the
standard deviation around this average). The proportion of
middle-income classes stays sensibly constant with respect to
population. In contrast, the order of the lower and higher income
fractions is inverted when the population crosses $2-300,000$
invidivuals.\label{fig:ratio_class}}
\end{figure*}

We plot these ratios on Fig.~\ref{fig:ratio_class} and we indeed see
that  the number of households belonging to the higher-income class is proportionally
larger in larger cities (for $H> 2-300,000$), while the number of households belonging to the
lower-income class is proportionally smaller. The proportion of Middle-income
class households stays essentially the same across all metropolitan areas.

In this work, we take a different approach and ask if the different
classes are more or less represented in a given MSA, compared to the
average US result. In this context, a city is richer if the
higher-income class is over-represented in this city, while the
lower-income class is under-represented. The measure stems from the
realisation that 'rich' and 'poor' are not absolute concepts, but must
be related to the environment.  In this case, it makes sense to
compare the representation of the different income classes
between metropolitan areas.

\section{Proportion of households in neighbourhoods}

\begin{figure}[!h]
    \centering
    \includegraphics[width=0.5\textwidth]{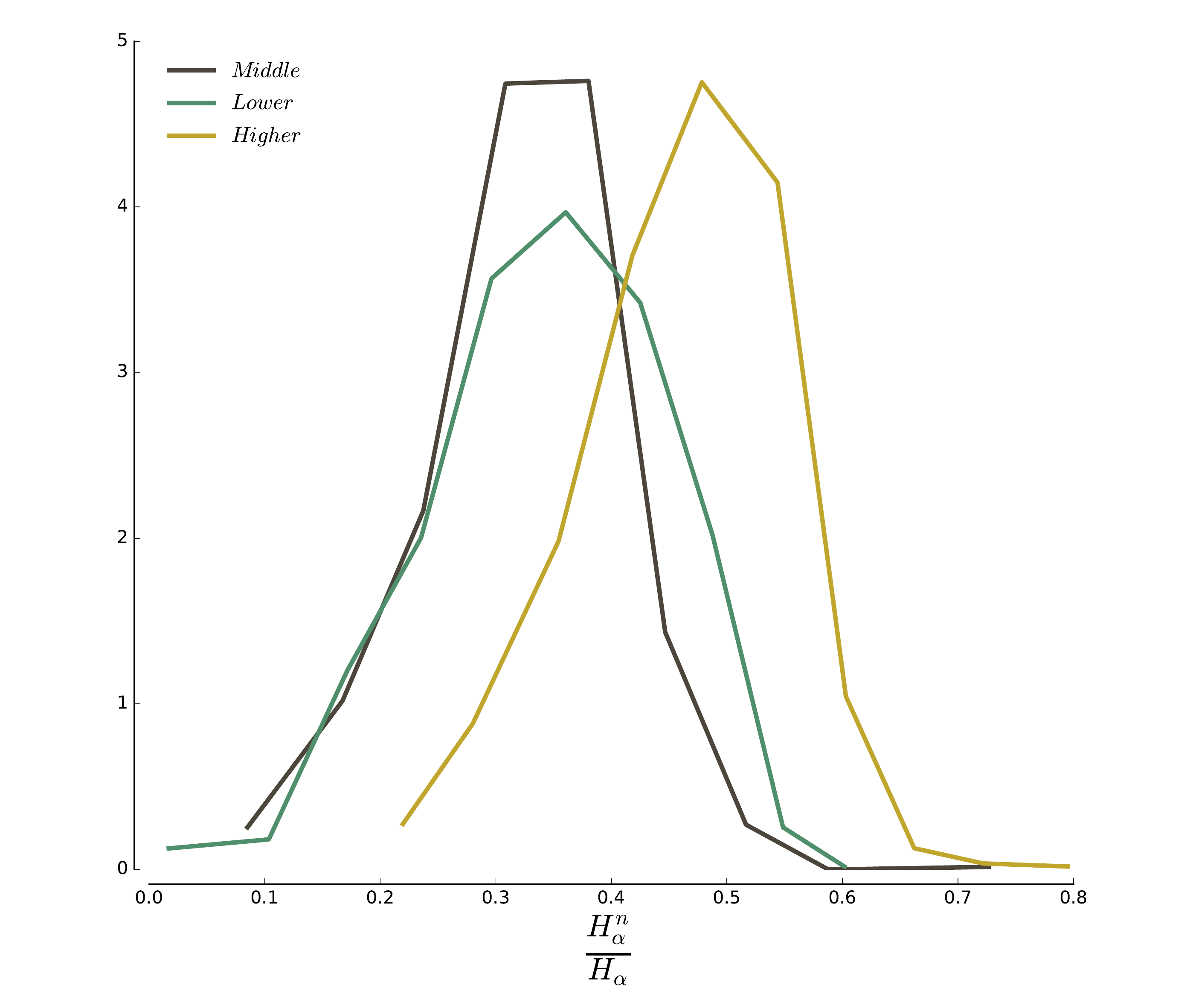}
    \caption{Distribution of the fraction of households belonging to a
      given class and that live in a neighborhood where it is
      over-represented (Middle, Lower, or Higher). \label{fig:ratio}}
\end{figure}

Neighbourhoods identify the areas in the city where the categories are
overrepresented, but this does not necessarily mean that most households
belonging to a category live in either of the corresponding neighourhood. We
plot the distribution of the proportion of households belonging to the lower-,
middle- and higher- income classes that also live in a corresponding
neighbourhood on Fig.~\ref{fig:ratio}.

One can see that higher-income households tend to be more concentrated in the
regions where they are represented, with an average of $52\%$. Followed by the
lower-income households, with and average of $40\%$. The middle-income
households are equally evenly spread across the city, 
with an average of $40\%$.

\section{Number of over-represented units and city size}

\begin{figure*}
    \centering
    \includegraphics[width=0.95\textwidth]{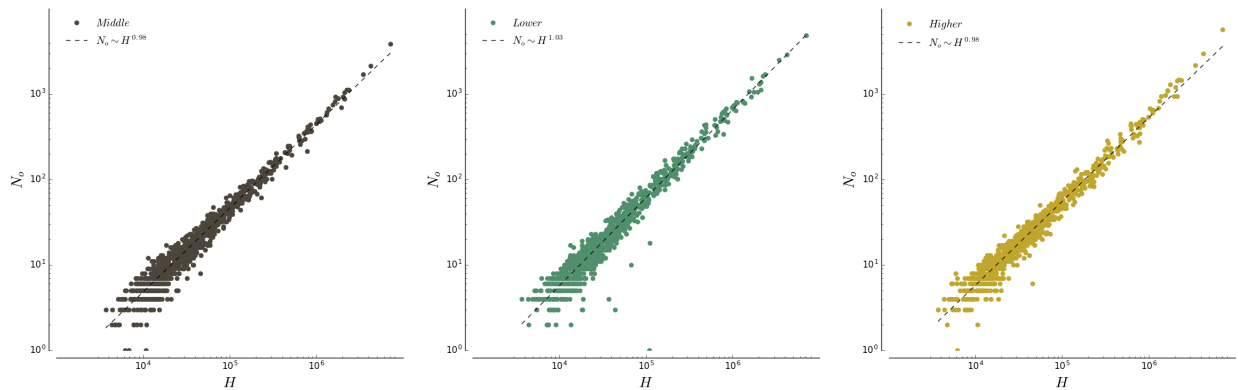}
    \caption{Number of areal units where each class is overrepresented as a
    function of the total number of households in the city. The behaviour is
consistent with a linear behaviour in the three cases.
\label{fig:overrepresented}}
\end{figure*}

In the main text, we find that the number of neighbourhoods for the $3$ classes
grows sublinearly with the size of a city, with a behaviour that is well
approximated by a power-law

\begin{equation}
    N_n \sim H^\nu
\end{equation}

with $\nu=0.86$ ($r^2 = 0.97$) for all classes together.
We claim this shows the tendency of classes to cluster more
in larger cities than in smaller ones. This is only true, however, if the number
of areal units in which each class is overrepresented does not itself vary
sublinearly with population size. We plot on Fig.~\ref{fig:overrepresented}
these numbers for each class and each city as a function of the size of the
city. We find that the behaviour of the number of overrepresented units is
consistent with a linear behaviour for all three classes

\begin{equation}
    N_o \sim H
\end{equation}

And our claim of increased clustering is thus justified.

\bibliographystyle{prsty}

\end{document}